\documentclass[11pt, a4paper]{article}
\usepackage[T1]{fontenc}
\usepackage{a4wide, url}
\usepackage[english]{babel}
\usepackage{graphicx}
\usepackage{natbib}
\usepackage[textfont=sc,labelfont=bf]{caption}
\usepackage[blocks]{authblk}
\usepackage{setspace}
\usepackage{multirow}
\usepackage{amsthm}
\usepackage{amsmath}
\usepackage{amsfonts}
\usepackage{amssymb}
\usepackage{pgf} 
\usepackage{bm}
\usepackage{paralist}
\usepackage{multirow}

\newcommand{\mbf}{\mathbf}
\newcommand{\mi}{\mathit}
\newcommand{\beq}{\begin{equation}}
\newcommand{\eeq}{\end{equation}}
\newcommand{\bea}{\begin{eqnarray}}
\newcommand{\eea}{\end{eqnarray}}
\newcommand{\ben}{\begin{enumerate}} 
\newcommand{\een}{\end{enumerate}}
\newcommand{\bpm}{\begin{pmatrix}}
\newcommand{\epm}{\end{pmatrix}}
\newcommand{\bbm}{\begin{bmatrix}}
\newcommand{\ebm}{\end{bmatrix}}

\newcommand{\ra}{\rightarrow}
\newcommand{\ba}{\begin{array}}  
\newcommand{\ea}{\end{array}}      
\newcommand{\bi}{\begin{itemize}}
\newcommand{\ei}{\end{itemize}}
\renewcommand{\r}{\right}
\renewcommand{\l}{\left}
\newcommand{\E}{\mathrm E}

\newcommand{\nn}{\nonumber}

\newtheorem{assumption}{Assumption}

\DeclareMathAlphabet\mathbfcal{OMS}{cmsy}{b}{n}

\begin{document}
\title {\textbf{\Large{Networks,  Dynamic Factors, and the Volatility Analysis\\ of High-Dimensional  Financial Series 
 }  }}
\author {{\sc Matteo Barigozzi$^{1}$}\hskip 1cm
{\sc Marc Hallin$^{2}$}
}
\date{}

\maketitle

\begin{abstract}
\noindent
We consider weighted directed networks for analysing, over the period 2000-2013, the interdependencies between volatilities of a large panel of stocks belonging to the S\&P100 index. In particular, we focus on the so-called {\it Long-Run Variance Decomposition Network}  (LVDN), where the nodes are stocks, and   the weight associated with edge $(i,j)$ represents the proportion of $h$-step-ahead forecast error variance of variable $i$ accounted for by variable $j$'s innovations. To overcome the curse of dimensionality, we decompose the panel into a component driven by few global, market-wide, factors,  and an idiosyncratic one modelled by means of a sparse vector autoregression (VAR) model. 
Inversion of the VAR together with suitable identification restrictions, produces the estimated network, by means of which we can assess how {\it systemic} each firm is.~Our analysis demonstrates the prominent role of financial firms as sources of contagion, especially during the~2007-2008 crisis. 
\end{abstract}

 \footnotetext[1]{m.barigozzi@lse.ac.uk -- London School of Economics and Political Science, Department of Statistics, UK.} 

\footnotetext[2]{mhallin@ulb.ac.be -- ECARES, Universit\'e libre de Bruxelles CP114/4 B-1050 Brussels, Belgium.\\

 }

\section{Introduction}\label{sec:intro}

The study of networks as complex systems has been the subject of intensive research in recent years, both in the physics and statistics communities \citep[see, for example,][for a review of the main models, methods, and results]{kolaczyk2009statistical}.~Typically, the datasets considered in that literature exhibit a ``natural'' or pre-specified network structure, as, for example, world trade fluxes \citep{serrano2003topology,barigozzi2010multinetwork}, co-authorship relations \citep{newman2001structure}, power grids \citep{watts1998collective}, social individual relationships \citep{zachary1977information}, fluxes of migrants \citep{fagiolo2014human}, or political weblog data \citep{adamic2005political}.
In all those studies, the network structure (as a collection of vertices and edges) is  known, and pre-exists  the observations. More recently, in the aftermath of the Great Financial Crisis of 2007-2008, networks also have become a  popular tool in financial econometrics and, more particularly, in the study of the interconnectedness of financial markets \citep{Diebold:Yilmaz:2013}. In this case, however, the data under study---usually time series of stock returns and volatilities---do not have any  particular pre-specified network structure,  and the graphical structure (viz., the collection of edges) of interest has to be recovered or estimated from the data. 


In this paper, we focus on one particular network structure: the Long-Run Variance Decomposition Network (LVDN). Following \citet{Diebold:Yilmaz:2013}, the LVDN, jointly with appropriate identification assumptions, defines, for a given horizon $h$, a weighted and directed graph where the weight associated with   edge $(i,j)$ represents the proportion of~$h$-step ahead forecast error variance of variable~$i$ which is accounted for by the innovations in variable~$j$. Therefore, by definition, LVDNs are completely characterised by the infinite vector moving average (VMA) representation entailed by  Wold's classical  representation theorem. 

Classical network-related quantities as {\it in} and {\it out node strength} or {\it centrality}, computed for a given LVDN, then admit an immediate economic interpretation in indicating, for instance,  which are the stocks or the firms most  affected by a global extreme event, and which are the stocks or the  firms that, when hit by some extreme shock, are likely to transmit it and spread it over to the whole market. The weights attached to each edge provide a quantitative assessment of the risks attached to such events. That type of network-based analysis is  of particular relevance in   financial econometrics---see, among others, \citet{Billio:Getmanksi:Lo:Pellizzon:2012}, \citet{acemoglu2013systemic}, \citet{hautsch2014forecasting,hautsch2014financial}, or \citet{Diebold:Yilmaz:2013}. 

Throughout, we are concentrating on the analysis of the LVDN associated with a panel of daily volatilities of the stocks constituting the Standard \&Poor's 100 (S\&P100) index. The observed period is from 3rd January 2000 to 30th September 2013. The stocks considered belong to ten different sectors: Consumer Discretionary, Consumer Staples, Energy, Financial, Health Care, Industrials, Information Technology, Materials, Telecommunication Services, Utilities. 

\begin{figure}[t]
	\begin{center}
	\begin{tabular}{cc}	
	\\
	
	{\includegraphics[width=0.45\textwidth]{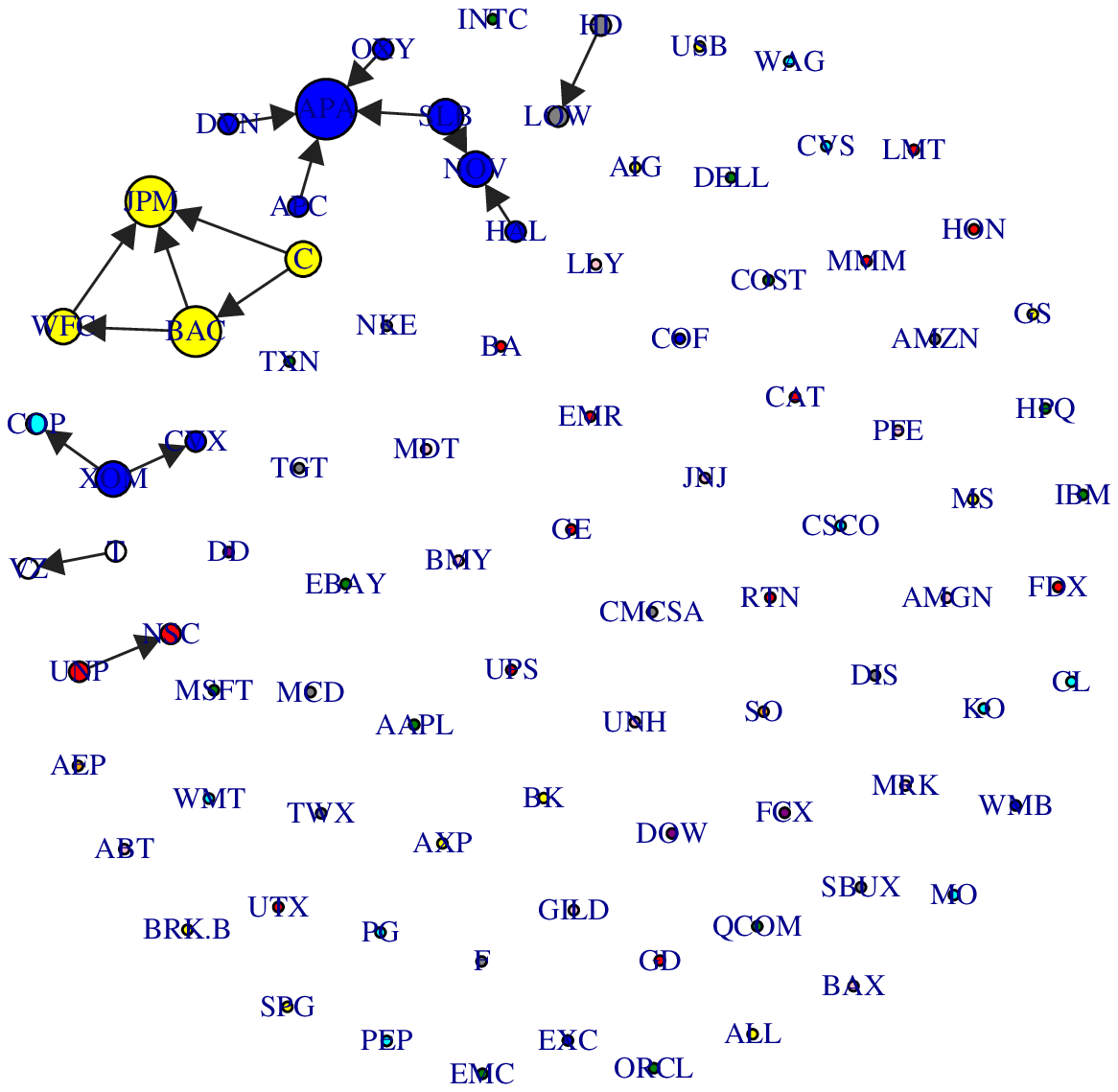}}	&	
	{\includegraphics[width=0.45\textwidth]{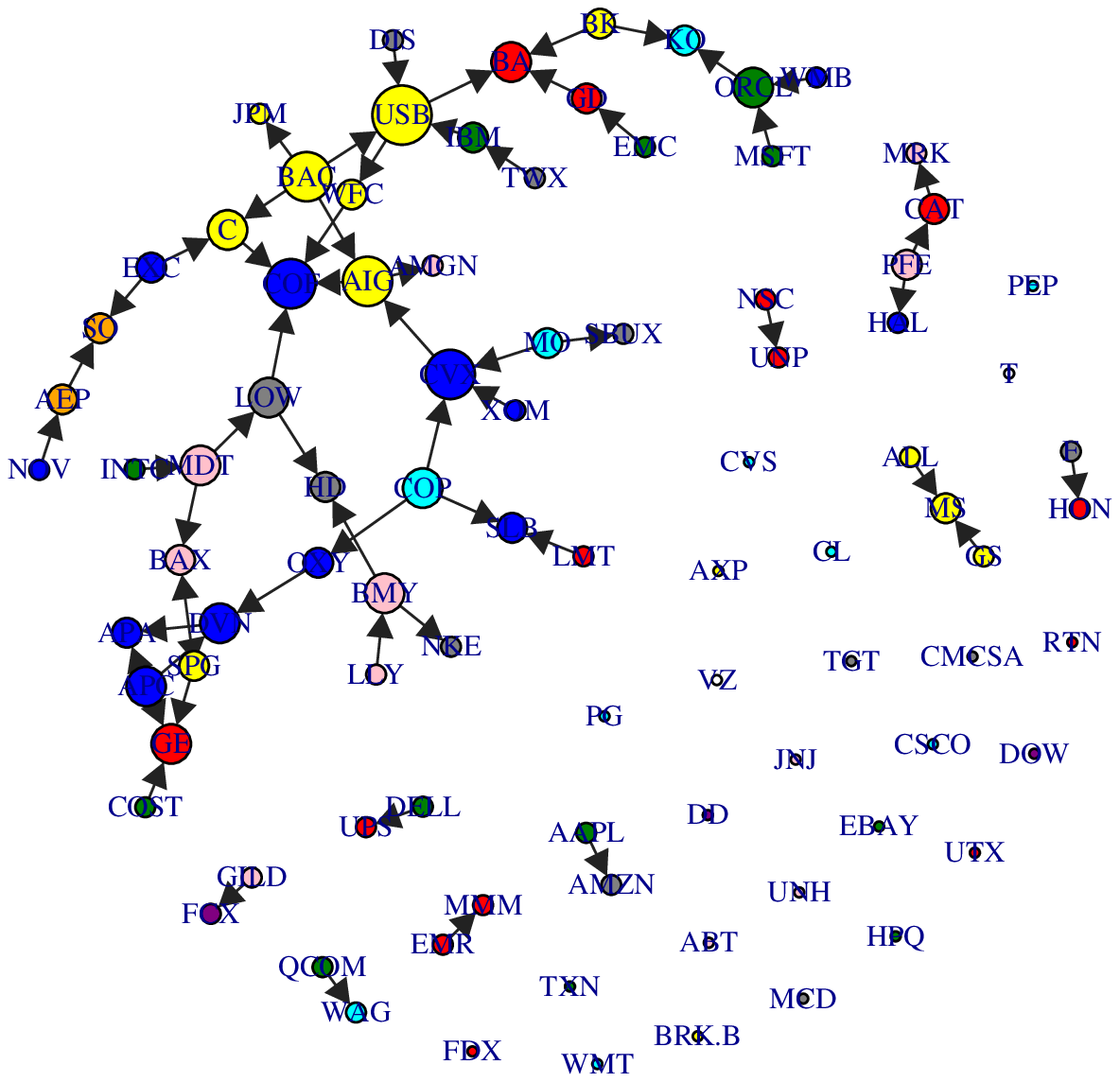}}\\
	\vspace{-7mm}\\
	\footnotesize{2000-2013} &\footnotesize{2007-2008}\\
	\end{tabular}
		\vspace{-2mm}
	\end{center}
		\caption{\textnormal{Graphs of estimated LVDN for the S\&P100 idiosyncratic volatilities.}} \label{fig:VDN_oo_thresh2}
\end{figure}

Our main findings are illustrated in Figure~\ref{fig:VDN_oo_thresh2} which shows, for our dataset, the estimated LVDNs associated (a) with the period 2000-2013, and (b) with  the years 2007-2008, which witnessed the so-called ``Great Financial Crisis''. 
Inspection of these graphs reveals a main role of the Energy (blue nodes) and Financial (yellow nodes) sectors. The interconnections within and between those two sectors had a prominent role in the period considered, due to the high energy prices in the years 2005-2007 and the Great Financial Crisis of the years~2007-2008.  In particular, when focussing only on the 2007-2008 period, the Financial stocks appear to be the most central ones \citep[in the sense of the eigenvector centrality concept of][]{bonacich2001eigenvector2}, and the connectedness of all network structures considered increases quite sizeably, making the whole system considerably more prone to contagion. This increased connectedness is an unsurprising phenomenon, since volatility measures {\it fear} or {\it lack of confidence} of investors, which tends  to  spread during periods of high uncertainty.
 
The LVDNs in Figure~\ref{fig:VDN_oo_thresh2} are estimated in two steps. First, we obtain what we call the \textit{idiosyncratic components} of volatilities by removing from the data the pervasive influence of global volatility shocks, to which we refer as \textit{common shocks} or, given the present financial context, as \textit{market shocks}. This is done by applying the general dynamic factor methodology recently proposed by \citet{FHLZ15a, FHLZ15b} and adapted by \citet{BH15} to a study of volatilities. More precisely, the factor model structure we are considering here is the Generalised Dynamic Factor model (GDFM) originally proposed by \citet{FHLR00} and  \citet{fornilippi01}. In a second step, the LVDN is obtained by estimating and inverting a sparse vector autoregression (VAR) for the resulting  idiosyncratic components, together with suitable identifying constraints. In particular, we consider VAR estimation based on three different methods: elastic net \citep{zou2005regularization}, group lasso \citep{yuan2006model}, and adaptive lasso \citep{zou2006adaptive}. We call this estimation approach ``factor plus sparse VAR'' approach.

This paper gives two main contributions to the existing financial literature on networks. First, we show that a combination of dynamic factor analysis and penalised regressions provides an ideal tool in the analysis of volatility and interconnectedness in large panels of financial time series. Second, we generalise to the high-dimensional setting the LVDN estimation originally proposed by \citet{Diebold:Yilmaz:2013} for a   small number of series. 
%

%
%

%
%
There are  strong  reasons in favour of  our approach controlling for market volatility shocks---as opposed to a direct sparse VAR analysis that does not control for those shocks. The main motivation is of an economic nature; but forecasting and empirical motivations are important as well. 

{\it (1) Economic motivation.} In the financial context,  the pertinence of factor models    is  a direct consequence of the Arbitrage Pricing Theory (APT) and the related Capital Asset Pricing Model (CAPM) \citep{ross1976arbitrage,FF93}. These models allow us to disentangle and identify the main sources of variation driving large panels of financial time series: {\it (i)} a strongly pervasive component typically driven by few common shocks, or factors, affecting the whole market, and {\it (ii)} an idiosyncratic weakly connected component driven by local, or sectoral, shocks. In agreement with APT, the factor-driven, or common, component represents the {\it non-diversifiable} or {\it systematic} part of risk, while the idiosyncratic component, becomes perfectly {\it diversifiable} as the dimension of the system grows \citep{chamberlainrotshild83}. Many studies, however, provide evidence  that   connectivity in the idiosyncratic component, although milder than in the common component,  still may be quite non-negligible, even in large dimensional systems \citep[see for example the empirical and theoretical results by][]{jovanovic1987micro,gabaix2011,acemoglu2012network}. This  is bound to happen in highly interconnected systems as financial markets. Therefore, when an exceptionally large shock, such as  bankruptcy,  affects the idiosyncratic component of a particular stock, that shock, although   idiosyncratic,  subsequently is likely to spread, and hit, eventually, all idiosyncratic components across the system. Such events are called {\it systemic}, and  diversification strategies  against them might be ineffective. Studying the LVDN related to the idiosyncratic volatilities, hence after controlling for market  shocks,  helps identifying the systemic elements of a panel of times series, and provides the basis for an analysis of contagion mechanisms.

{\it (2) Forecasting motivation.} It has been shown \citep[see e.g.][]{de2008forecasting} that forecasts obtained via penalised regression are highly unstable in the presence of collinearity. Thus, even though forecasting is not the main goal of this paper, removing the effect of common shocks before turning to sparse VAR estimation  methods seems highly advisable. 

{\it (3) Empirical motivation.} When considering partial dependencies, measured by partial spectral coherence, in the idiosyncratic component, that is after common components have been removed, hidden dependencies between and within the Financial and Energy sectors are uncovered. This finding, documented in Section~\ref{sec:emp}, is the empirical justification for preferring a ``factor plus sparse VAR'' approach rather than a  direct application of sparse VAR methods.

In Section \ref{sec:gdfm}, we introduce the GDFM for large panels of time series and the definition of LVDN for the idiosyncratic components, allowing for a study of different sources of interdependencies. In Section \ref{sec:est} we discuss estimation. In Section \ref{sec:vol}, following \citet{BH15}, we show how to extract from financial returns those volatility proxies which will be the object of our analysis. Section~\ref{sec:emp} presents the results for the S\&P100 dataset. 
A detailed list of the series considered and complementary results are provided in the Web-based supporting materials.

\section{Factors and networks in large panels of financial time series }\label{sec:gdfm}

We consider large panels of time series data, namely, observed  finite realizations, of the form~$\{Y_{it}\vert \, i=1,\ldots ,n,\ t=1,\ldots , T\}$, of some stochastic process  $\mbf Y:=\{Y_{it}\vert\,  i\in\mathbb{N},\ t\in~\!\mathbb{Z}\}$;  $i$ is a cross-sectional index and $t$ stands for time.  Both the cross-sectional dimension $n$ and the sample size or  series length $T$ are large and, in asymptotics, we consider sequences of $n$ and $T$ values tending to infinity. The notation $\mbf Y_n:= \{\mbf Y_{nt}=(Y_{1t}, Y_{2t},\ldots ,Y_{nt})'\vert\, t \in \mathbb Z\}$ in the sequel is used  for the $n$-dimensional subprocess of 
  $\mbf Y$; 
 the same notation is used for all $n$-dimensional vectors. In this section, $\mbf Y_n$ stands for  a generic panel of time series. Since our interest is to study connections and  interdependencies responsible for the contagion phenomenons that might lead to  financial crises, then  in Sections~\ref{sec:vol} and~\ref{sec:emp}, we will apply the definitions and results presented  here to the case of financial volatilities.
    
In principle, as shown by \citet{Diebold:Yilmaz:2013}, a LVDN can be  estimated via  classical VAR estimation. However, when dealing with high-dimensional systems, VAR estimation is badly affected by curse of dimensionality problems, and  adequate estimation techniques have to be  considered. The  most frequent strategy, in presence of large-$n$ datasets, is  based on sparsity assumptions allowing for the application of penalised regression techniques \citep{hsu2008subset,abegaz2013sparse,nicholson2014hierarchical,basu2015,Davis:Zang:Zheng:2012,Kock:Callot:2012,nets14,gelper2015identifying}. The presence of pervasive shocks affecting  the large panels of time series considered in macro-econometrics and finance has motivated the development of another dimension reduction technique, the so-called dynamic factor methods. Various versions of those methods have become daily practice in many areas of econometrics and finance; among these the GDFM by \citet{FHLR00} and \citet{fornilippi01} is the most general, while most other factor models considered in the time series literature  \citep[][to quote only a few]{stockwatson02JASA,baing02,LY12,FLM13} are particular cases.

\subsection{The Generalised Dynamic Factor Model}\label{DFsec}
In order to introduce the {\it dynamic factor representation} for $\mbf Y$, we make the following assumptions.
\begin{assumption}
For all $n \in\mathbb N$, the vector process $\mbf Y_n$ is second-order stationary,  with mean zero and  finite variances.\end{assumption}

\begin{assumption} For all $n \in\mathbb N$, the spectral measure of  $\mbf Y_n$ is absolutely
continuous with respect to the Lebesgue measure on $[-\pi, \pi]$, that is, $\mbf Y_{n}$ admits 
a full-rank (for any $n$ and~$\theta$)
 spectral density matrix $\mi\Sigma_{\mbf Y;n}(\theta)$, $\theta\in[-\pi,\pi]$ with   uniformly (in $i$, $j$, $\theta$, and $n$) bounded  entries~$\sigma_{{\bf Y};ij}(\theta )$. \end{assumption}
 
Denote by   $\lambda_{\mbf Y;n,1}(\theta),\ldots , \lambda_{\mbf Y;n,n}(\theta)$, $\theta \in [-\pi, \pi]$,  the eigenvalues (in decreasing order of magnitude) of  $\mi\Sigma_{\mbf Y;n} (\theta)$;  the mapping $\theta\mapsto \lambda_{\mbf Y;n,i}(\theta)$ is also called 
$\mbf Y_n$'s $i$th {\it dynamic eigenvalue}. We say that $\mbf Y$ admits a General Dynamic Factor representation with $q$ factors if for all $i$ the process $\{Y_{it}\}$ decomposes into a {\it common} component $\{X _{it}\}$ and an {\it idiosyncratic} one  $\{Z _{it}\}$,
\begin{eqnarray}
Y_{it} = X_{it} + Z_{it} =:\sum_{k=1}^q b_{ik} (L)u_{kt} + Z_{it}, \quad i\in\mathbb N,\quad t\in\mathbb Z,\label{eq:gdfm_level}
\end{eqnarray}
such that
  \medskip
\begin{compactenum}
\item [({\it i})] the $q$-dimensional vector process of factors $\mbf u:=\{\mbf u_t= (u_{1t}\; u_{2t}\ldots u_{qt} )'\vert\,t\in\mathbb Z\}$ is  orthonormal zero-mean white noise;
\item [({\it ii})] the filters $b_{ik} (L)$ are one-sided and square-summable for all
$i\in\mathbb N$ and $k=1,\ldots,q$;
\item [({\it iii})] the $q$th dynamic eigenvalue $\lambda_{\mbf X;n,q}(\theta)$ of  
 $\mbf X_n$ 
 diverges  $\theta$-almost everywhere ($\theta$-a.e.) in the interval~$[-\pi,\pi]$ as~$n\to\infty$;
\item [({\it iv})] the first dynamic eigenvalue $\lambda_{\mbf Z;n,1}(\theta)$ of 
 $\mbf Z_n$ 
   is  bounded ($\theta$-a.e.~in $[-\pi,\pi]$) as $n\to\infty$; 
\item [({\it v})] $Z_{k,t_1}$ and $u_{h,t_2}$ are mutually orthogonal for any $k$, $h$, $t_1$ and $t_2$;
\item [({\it vi})] $q$ is minimal with respect to ({\it i})-({\it v}).
\end{compactenum}
\medskip
For any $n$, we can write \eqref{eq:gdfm_level} in vector notation as
\begin{equation}\label{eq:gdfm_level'}
\mbf Y_{nt} =  \mbf X_{nt} +  \mbf Z_{nt}=: \mi B_n(L) \mbf u_t + \mbf Z_{nt},\quad n\in\mathbb N,\quad t\in \mathbb Z.
\end{equation}
This actually defines the GDFM. 
In this model, the common and idiosyncratic components are identified by means of the following assumption on $\mbf Y_n$'s dynamic eigenvalues. 

\begin{assumption}
The $q$th   eigenvalue $\lambda_{\mbf Y;n,q}(\theta)$ of $\mi\Sigma_{\mbf Y;n} (\theta)$  diverges, $\theta$-a.e. in $[-\pi,\pi]$, while the~$(q+1)$th one, $\lambda_{\mbf Y;n,q+1}(\theta)$, is $\theta$-a.e. bounded, as~$n\ra\infty$. 
\end{assumption}

More precisely, we know from \citet{FHLR00} and \citet{fornilippi01} that, given Assumptions~1 and 2, Assumption 3 is necessary and sufficient for the process $\mbf Y$ to admit the dynamic factor representation \eqref{eq:gdfm_level}. \citet{HL13} moreover provide very weak time-domain primitive conditions under which \eqref{eq:gdfm_level}, hence Assumption 3, holds for some $q<\infty$. 

Finally, the idiosyncratic component $\mbf Z_n$ always admits a Wold decomposition which, after  adequate transformation, yields the vector moving average (VMA) representation 
\beq\label{eq:woldrep2}
\mbf Z_{nt} := \mi D_n(L)\mbf e_{nt},\qquad 
 t\in\mathbb Z, \quad  \mbf e_{nt} \sim w.n.({\bf 0},{\mi I}_n),
\eeq
where $\mi D_n(L)=\sum_{k=0}^{\infty} \mi D_{nk}L^k$ is a square-summable power series in the lag operator $L$. Notice that although the magnitude of those coefficients is bounded by condition (iv) above, no sparsity assumption is made. 

\subsection{The Long-Run Variance Decomposition Network}\label{LVDNsec} 
In order to study the interdependencies among series, and following traditional econometric analysis, we focus on the reactions of observed variables to unobserved shocks, i.e.~impulse response functions. Large panels of financial time series are affected by market shocks that are, essentially,  common to all stocks and represent the non-diversifiable components of risk,  and by idiosyncratic shocks that are specific to one or a few stocks in the panel.  The GDFM   is the ideal tool for disentangling those two sources of variation: (i) the $q$ market shocks~$\mbf u$ and their impulse responses $\mi B_n(L)$, defined in \eqref{eq:gdfm_level'}, and (ii) the $n$ idiosyncratic shocks~$\mbf e_n$ and their impulse responses $\mi D_n(L)$, defined in \eqref{eq:woldrep2}. 

Our focus  here is mainly on idiosyncratic shocks. Indeed, once we control for market effects, the study of interdependencies among different stocks is strictly related to systemic risk measures, that is,  individual measures of how one given stock is likely to be affected by,  and/or is likely to affect, all others. 

To this end, representation \eqref{eq:woldrep2} is what we need,  as it characterises all   (linear)  interdependencies between the components of~$\mbf Z_{n}$. In particular,~$\mi D_{n0}$ characterises contemporaneous dependencies, while $\mi D_{nk}$ for~$k>0$ characterises the dynamic dependencies with lag $k$. Denote by $d_{k,ij}$ the $(i,j)$th entry of~$\mi D_{nk}$. Then, following \citet{Diebold:Yilmaz:2013}, we can summarise all  dependencies up to  lag $h$ by means of the forecast error variance decomposition and, more particularly, by the  ratios 
\beq\label{eq:dijh}
w_{ij}^h := 100\l(\frac{\sum_{k=0}^{h-1} d_{k,ij}^2 }{\sum_{\ell =1}^n \sum_{k=0}^{h-1} d_{k,i\ell }^2 }\r),\quad i,j = 1,\ldots ,n.
\eeq
The ratio $w_{ij}^h$  is the  percentage of the $h$-step ahead forecast error variance of $\{Y_{it}\}$  accounted for by the innovations in $\{Y_{jt}\}$. Note that, by definition, 
\[
\frac 1 {100}\sum_{j=1}^n w_{ij}^h=1\;\text{ for any $i$, hence }\; \frac 1 {100}\sum_{i,j=1}^n w_{ij}^h=n.
\]
The LVDN is then defined by the set of edges 
\beq
\mathcal E_{\text{\tiny LVDN}}:=\Big\{(i,j)\in\{1\ldots n\}^2 \vert\ w_{ij}^{\text{\tiny LVDN}}:=\lim_{h\to\infty} w_{ij}^{h}\neq 0\Big\}.
\eeq
In practice an horizon $h$ has to be chosen to compute those weights, and the operational  definition of the LVDN  therefore also  depends on that $h$.

Three   measures of connectedness can be based on the quantities defined in~\eqref{eq:dijh}. First,  we define the {\it from-degree} of component $i$ and  {\it to-degree} of component  $j$ (also called   {\it in-strength} of node $i$  and {\it out-strength} of node  $j$)   as 
\begin{equation}\label{eq:indeg}
\delta^{\mbox{\tiny{From}}}_{i} := \sum_{\substack{j=1,{j\neq i}}}^n w_{ij}^{h},\quad i=1,\ldots, n \quad\text{and}\quad 
\delta^{\mbox{\tiny{To}}}_{j} := \sum_{\substack{i=1,{i\neq j}}}^n w_{ij}^{h},\quad j=1,\ldots, n, 
\end{equation}
respectively. As pointed out by \citet{Diebold:Yilmaz:2013}, these two measures are closely related to two classical measures of systemic risk considered in the financial literature.   The from-degree  is  directly  related with the so-called {\it marginal expected shortfall} and {\it expected capital shortfall} (of series~$i$),  which measure the exposure of component $i$ to extreme events affecting all other components \citep[see][for a definition of these measures]{acharya2012}. As for the  to-degree,
  it is related to {\it co-Value-at-Risk}, which measures the effect on the whole panel of an extreme event affecting component~$j$ \citep[see][for a definition]{adrian2011}.
Finally, we can define an overall measure of connectedness by summing all from-degrees (equivalently,  all to-degrees):
\beq\label{eq:totdeg}
\delta^{\mbox{\tiny{Tot}}}:=\frac 1 n\sum_{i=1}^n\delta^{\mbox{\tiny{From}}}_{i}=\frac 1 n\sum_{j=1}^n\delta^{\mbox{\tiny{To}}}_{j}.
\eeq
Given the economic interpretation of these quantities, the LVDN of the  idiosyncratic component $\mbf Z_n$ seems to be an ideal tool for studying systemic risk and, for this reason, in the empirical study of Section \ref{sec:emp}, we mainly focus on the LVDN of volatilities (see also Section \ref{sec:vol} for a motivation).

An LVDN also can be constructed  for the common component $\mbf X_n$ by using a definition analogous to \eqref{eq:dijh}, but based on the entries of the matrix polynomial $\mi B_n(L)$, defined in \eqref{eq:gdfm_level'}. However, due to the singularity of $\mi B_n(L)$,  definition~\eqref{eq:dijh} in this case does not measure the proportion of the $h$-step ahead forecast error variance of variable $i$ accounted for by the innovations in variable~$j$, but rather the proportion of the same  forecast error variance explained by the $j$th market shock $\{u_{jt}\}$.

\subsection{VAR representations}
The LVDN of the idiosyncratic component $\mbf Z_n$ is defined from the coefficients of the VMA representation (\ref{eq:woldrep2}). That representation can be estimated as an inverted sparse VAR. We accordingly   make the following assumption. 
\begin{assumption}
The idiosyncratic component $\mbf Z_n$ admits, for some $p$ that does not depend on~$n$,  the VAR($p$) representation
\beq\label{eq:varpidio}
\mi F_n(L) \mbf Z_{nt} = \mbf v_{nt},\quad t\in\mathbb Z,\quad \mbf v_{nt}\sim w.n.(\mbf 0,\mi C^{-1}_n),
\eeq
where $\mi F_n(L)=\sum_{k=0}^p \mi F_{nk}L^k$ with $\mi F_{n0}=\mi I_n$ and $\det(\mi F_n(z))\neq 0$ for any $z\in\mathbb C$ such that~$|z|\leq 1$, and $C_n$ has full rank. 
 Moreover, denoting by  $f_{k,ij}$ and $c_{ij}$
   the $(i,j)$th entries of $\mi F_{nk}$ and $\mi C_n$, 
\begin{align}
\max_{j=1\ldots n}\sum_{i=1}^n \mathbb I_{(f_{k,ij}\neq 0)} &= o(n), \quad k=1,\ldots,p,\quad  n\in\mathbb N,\label{eq:varcoeff}\\
\max_{i=1\ldots n}\sum_{j=1}^n \mathbb I_{(f_{k,ij}\neq 0)} &= o(n), \quad k=1,\ldots,p,\quad  n\in\mathbb N,\label{eq:varcoeff2}\\
\max_{j=1\ldots n}\sum_{i=1}^n \mathbb I_{(c_{ij}\neq 0)} &= o(n), \quad  n\in\mathbb N.\label{eq:ccoeff}
\end{align}
\end{assumption}
The first part  \eqref{eq:varpidio} of this assumption is quite mild, provided that $p$ can be chosen large enough. The second part requires some further clarification. 
In \eqref{eq:varcoeff}-\eqref{eq:varcoeff2}, we require the VAR coefficient matrices in  \eqref{eq:varpidio} to have only a small number of non-zero entries.
 In this sense we say that the VAR representation \eqref{eq:varpidio} is sparse \citep[see, for example, the definitions of sparsity in][]{Bickel:Levina:2008,Cai:Liu:2011}. That assumption is needed for the consistent estimation of \eqref{eq:varpidio} in the large-$n$ setting, and it is justified by the idea that in a GDFM, once we control for common shocks, most interdependencies among the elements of $\mbf Z_n$ are quite  weak (since the corresponding  dynamic eigenvalues are bounded as $n\to\infty$).  However, note  that, while a sparse VAR is related to conditional second moments, the GDFM assumptions on the idiosyncratic component are based on unconditional second moments. For this reason,    the GDFM assumptions do not imply  a sparse VAR representation for $\mbf Z_n$, and   \eqref{eq:varcoeff}-\eqref{eq:varcoeff2} are needed. Finally, for convenience, we parametrise the covariance matrix of the VAR innovations by means of its inverse $\mi C_n$ and in \eqref{eq:ccoeff} we require this matrix to be sparse too, in accordance  with the idea of a sparse global  conditional dependence structure of idiosyncratic components. 
 
As a by-product,  a Long-Run Granger Causality Network  (LGCN) can be defined by the set of edges
 \beq
\mathcal E_{\text{\tiny LGCN}}:=\Big\{(i,j)\in\{1\ldots n\}^2 \vert\  w_{ij}^{\text{\tiny LGCN}}:= \sum_{k=0}^p f_{k,ij}\neq 0\Big\}.
 \eeq
 This network  captures the leading/lagging conditional dependencies of a given panel of time series. Such graphical representations  of VAR dependencies were initially proposed by \citet{DE03} and \citet{Eichler:2007}, and  extend to a time-series context   the graphical models for independent data considered by \citet{Dempster:1972}, \citet{Meinshausen:Buhlmann:2006},  \citet{glasso}, and \citet{Peng:Wang:Zhou:Zhi:2009}, to quote only a few. 
 
Two comments are in order here. A network is said to be sparse if its weight matrix has many zero entries.  First, notice that, under Assumption 4,  the LGCN is likely to be sparse. On the other hand, the GDFM assumptions do not guarantee sparsity of the LVDN but only some weaker restrictions on the magnitude of its entries, as dictated by the boundedness of the eigenvalues of $\mbf Z_n$'s spectral density matrix. Second, the economic interpretation of the  LGCN  is not as straightforward as that of  the LVDN, and  the LGCN therefore is of lesser interest for the analysis of financial systems: mainly, it will be  a convenient tool in the derivation of the LVDN. This is in line with traditional macroeconomic analysis where impulse response functions, i.e. VMA coefficients, rather than VAR ones, are the object of interest for policy makers.

As for the common component $\mbf X_n$, \citet{FHLZ15a} show that it admits   the singular VAR representation 
\beq\label{eq:ARX}
\mi A_n(L)\mbf X_{nt} = \mi H_n\mbf u_t, \quad t\in\mathbb Z,\quad \mbf u_t\sim w.n.(\mbf 0,\mi I_q). 
\eeq 
Assuming, without loss of generality, that   $n=m(q+1)$ for some integer $m$, the VAR operator~$\mi A_n(L)$  in \eqref{eq:ARX}  is  block-diagonal,  with~$(q+1)\times (q+1)$-dimensional diagonal  blocks of the form~$\mi A^{(i)}(L)=\sum_{k=0}^{p_i} \mi A_{k}^{(i)}L^k$  such that, for any $i=1,\ldots, m$,  $\mi A_{0}^{(i)}=\mi I_n$ and $\det(\mi A^{(i)}(z))\neq 0$ for any $z\in\mathbb C$ such that $|z|\leq 1$. Moreover, $\mi H_n$ is a full-rank $n\times q$ matrix, and $\mbf u$ is the $q$-dimensional process of common shocks defined in \eqref{eq:gdfm_level'}.

\subsection{Identification}
Starting from $\mbf Z_n$'s VAR representation  \eqref{eq:varpidio} and  comparing it with \eqref{eq:woldrep2}, we have
\beq\label{eq:LVDNidio00}
\mi D_n(L)=(\mi F_n(L))^{-1}\mi R_n, 
\eeq
 where the full-rank matrix $\mi R_n$ is   making the shocks~$\mi R_n^{-1}\mbf v_{n}=:\mbf e_{n}$ orthonormal (such matrices under Assumption~4 exist). In other words, the LVDN is obtained from the inversion of the VAR in \eqref{eq:varpidio}   by selecting a suitable transformation $\mi R_n$ meeting  the required identification constraints. The simplest choice for $\mi R_n$ follows from   a Choleski decomposition of the covariance~$\mi C_n^{-1}$ of the  shocks  \citep[see][]{sims80}---namely,  selecting $\mi R_n$ as the lower triangular matrix  such that $\mi C_n^{-1}=\mi R_n^{}\mi R_n'$. Such a choice is appealing, as it is purely data-driven, but it depends on the ordering of the variables or, equivalently, on the ordering of the components of the  shocks vector $\mbf v_{n}$.  
 
Many   orderings  are possible, and  the one we propose  is  based on $\mbf v_{n}$'s partial correlation structure. More precisely, assuming $\mi C_n$ to have full rank, the partial correlation between $\{v_{it}\}$ and $\{v_{jt}\}$ is  
\beq\label{PCN1}
\rho^{ij}:=\frac{-c_{ij}}{\sqrt{c_{ii}c_{jj}}},\qquad i,j=1,\ldots, n
\eeq
where $c_{ij}$ is the $(i,j)$th entry of $\mi C_n$. Associated with this concept of partial correlation is the Partial Correlation Network (PCN), with edges
\beq\label{PCN2}
\mathcal E_{\text{\tiny PCN}}:=\{(i,j)\in\{1\ldots n\}^2 \vert\  w_{ij}^{\text{\tiny PCN}}:=\rho^{ij}\neq 0\},
\eeq
which, by Assumption 4,  is a sparse network. The PCN can be studied by means of the $n$ linear regressions
\beq\label{PCN3}
v_{it} =\sum_{j=1,j\neq i}^n \beta_{ij} v_{jt} +\nu_{it}, \quad t\in\mathbb Z,\quad \nu_{it}\sim w.n.(0,\sigma^2_i),\quad i=1,\ldots n.
\eeq
Indeed, it can be shown that $\beta_{ij}=\rho^{ij}\sqrt{\sigma_i^2/\sigma_j^2}$ \citep[see, for example, Lemma 1 in][]{Peng:Wang:Zhou:Zhi:2009}, so that $(i, j)\in \mathcal E_{\text{\tiny PCN}}$ if and only if $\beta_{ij}\neq 0$. 
 Thus, the inverse covariance matrix $\mi C_n$ of the VAR shocks is directly related to the partial correlation matrix of the VAR innovations, and this matrix in turn can be seen as the PCN weight matrix. We then order the shocks by decreasing order of  eigenvector centrality \citep[as in][]{bonacich1987power} in that PCN, the most central component receiving label  one, etc. 

The centrality measure considered defines each node's centrality as the sum of the centrality values of its neighbouring nodes. It is easily seen that the mathematical translation of this idea leads to an eigenvector-related concept of centrality. That concept of eigenvector centrality differs from that of degree centrality (the number of neighbours of a given node); indeed, a node receiving many links does not necessarily have high eigenvector centrality,  while, a node with high eigenvector centrality is not necessarily highly linked (it might have few but important linkers).

It has to be noticed that usually eigenvector centrality is defined for networks in which the sign of the weight associated to a given edge is not taken into account, so negative and positive partial correlations have the same importance. However, based on the idea that contagion is more likely between nodes which are linked through positive weights rather than negative ones, we can also consider eigenvector centrality for a signed network, thus preserving the information about weights' signs. Both approaches are considered in the empirical analysis that follows.

This identification strategy seems well suited to the study of financial contagion. Indeed, in an impulse response exercise, we study the propagation of shocks through the system starting from lag zero up to a given lag $h>0$. Thus, a given order of shocks defines which component we choose to hit first. By ordering nodes according the their centrality in the PCN of VAR residuals, we are considering the case in which the most contemporaneously interconnected node is firstly affected by an unexpected shock, and then, by means of the subsequent impulse response analysis, we study the propagation of such shock through the whole system. The corresponding row of the LVDN adjacency matrix gives a summary, in terms of explained variance, of the effect of this propagation mechanism after $h$ lags. 

Finally, it is worth mentioning a few other possible methods for identifying the shocks in the LVDN. One could rank the series based on endogenous characteristics of the firms considered, such as market capitalisation. Another data-driven approach is taken in \citet{swanson1997impulse}, who propose to test the over-identifying restrictions implied by orthogonal shocks. That approach is closely related to ours as those restrictions involve partial correlations of the shocks; however as the dimension $n$ of the problem increases, the related computational problems rapidly become non-tractable. \citet{Diebold:Yilmaz:2013} adopt generalised variance decompositions, a very popular method originally proposed by \citet{koop1996impulse}. This approach, however, is only valid if the shocks have a Gaussian distribution, an assumption which is unlikely to hold true for the S\&P100 dataset.

Turning to the common component $\mbf X_n$, its LVDN is obtained by inverting the VAR representation~\eqref{eq:ARX}. Namely, there exists a $q\times q$ invertible matrix $\mi K$ such that
\beq\label{eq:LVDNcom4}
\mi B_n(L) = (\mi A_n(L))^{-1}\mi H_n\mi K. 
\eeq
Here $\mi K$ is required in order to identify the orthonormal market shocks $\mbf u$.  
However,  if~$q=1$ as in the empirical analysis of Section~\ref{sec:emp}, the choice of $\mi K$ reduces to that of a sign and a scale. 

In the next sections, we first discuss the estimation of  GDFMs and  LVDNs under the general definition of this section, and then  adapt   to the particular case of unobserved volatilities. 


\section{Estimation}\label{sec:est}

In this section we review estimation of \eqref{eq:LVDNidio00} and \eqref{eq:LVDNcom4}. The numbers of factors throughout are  determined via the information criterion proposed by \citet{hallinliska07} and based on the behaviour of dynamic eigenvalues as the panel size grows from 1 to $n$. In this section we consider a generic observed panel $
\{Y_{it} \vert\ i=1,\ldots,n,  \  t=1,\ldots, T\}$ of time series, with sample size $T$ (in the next section these would be either returns or volatilities). Hereafter, we use the superscript $T$ to denote estimated quantities.

\subsection{GDFM estimation}\label{sec:est1}
First we recover the common component using its autoregressive representation \eqref{eq:ARX}. For a given number of factors $q$, the method, described in detail by \citet{FHLZ15a,FHLZ15b}, is based on the following steps.\medskip
\begin{compactenum}
\item [\textit{(i)}] Estimate the spectral density matrix of $\mbf Y_n$, denoted as $\mi\Sigma_{\mbf Y;n}^T(\theta)$, for example using the smoothed periodogram estimator.
\item [\textit{(ii)}] Use the $q$ largest dynamic principal components of $\mi\Sigma_{\mbf Y;n}^T(\theta)$ to extract the spectral density matrix of $\mbf X_n$, denoted as $\mi\Sigma_{\mbf X;n}^T(\theta)$ \citep[see][]{brillinger1981}.
\item [\textit{(iii)}] Compute the autocovariances of $\mbf X_n$ by inverse Fourier transform of $\mi\Sigma_{\mbf X;n}^T(\theta)$ and use these to compute the Yule-Walker estimator of the VAR filters $\mi A_n^T(L)$; denote by $\bm\epsilon_{n}^T:=\{\epsilon_{it}^T\vert\ i=1,\ldots,n,  \  t=1,\ldots, T\}$ the corresponding residuals.
\item [\textit{(iv)}] From the sample covariance of $\bm\epsilon_{n}^T$, compute the eigenvectors corresponding to its $q$ largest eigenvalues, these are the columns of the estimator $\mi H_n^T$;  then, by projecting $\bm\epsilon_{n}^T$ onto the space spanned by the columns of $\mi H_n^T$, obtain the $q$-dimensional vector $\{\mbf u_{t}^T\vert\ t=1,\ldots, T\}$.
\item [\textit{(v)}] The estimated LVDN of the common component is given by $(\mi A_n^T(L))^{-1}\mi H_n^T\mi K$, where $\mi K$ is a generic $q\times q$ invertible matrix such that $\mi H_n^T\mi K$   consistently estimates $\mi H_n$.
\item [\textit{(vi)}]  The estimated common and idiosyncratic components are respectively given by
$$
\mbf X_{nt}^T=(\mi A_n^T(L))^{-1}\mi H_n^T\mbf u_t^T \quad\text{and}\quad   \mbf Z_{nt}^T=\mbf Y_{nt}-\mbf X_{nt}^T, \quad t=1,\ldots ,T.
$$
\end{compactenum}
Details on each step and asymptotic properties of the estimators are given in \citet{FHLZ15b}. In particular, the parameters of the model are estimated consistently as $n$ and $T$ tend to infinity,  at $O_P(\max({n^{-1/2},T^{-1/2}}))$ rate.

\subsection{Sparse VAR estimation}\label{sec:est2}
Once we have an estimator of the $n$ idiosyncratic components, we estimate  the VAR($p$) representation (\ref{eq:varpidio})   
by minimising the penalised quadratic loss
\beq\label{eq:loss}
\mathcal L_T=  \sum_{t=1}^T \bigg( Z_{it}^T-\sum_{k=1}^p \mbf f_{k,i}' \mbf Z_{nt-k}^T\bigg)^2+\mathcal P(\mbf f_{1,i} \ldots, \mbf f_{p,i}),\quad i=1,\ldots, n,
\eeq
where $\mbf f_{k,i}'$ is the $i$-th row of $\mi F_{kn}$ and $\mathcal P(\cdot)$ is some given penalty which depends on the chosen estimation method. In particular, we consider three alternative strategies.
\medskip
\begin{compactenum}
\item [\textit{(i)}] Elastic net, as defined by \citet{zou2005regularization}, where the penalty   is a weighted average of ridge and lasso penalties:
\[
\mathcal P^{EN}(\mbf f_{1,i} \ldots, \mbf f_{p,i}) = \lambda \Vert (\mbf f_{1,i}' \ldots \mbf f_{p,i}')'\Vert_1 + (1-\lambda)\Vert (\mbf f_{1,i}' \ldots \mbf f_{p,i}')'\Vert_2^2.
\]
\item [\textit{(ii)}] Adaptive lasso, as defined by \citet{zou2006adaptive}, where the penalty is a lasso one but conditioned on a pre-estimators $\widetilde{\mbf f}_{k,i}$ of the parameters (typically given by ridge or least squares estimators)
\[
\mathcal P^{AL}(\mbf f_{1,i} \ldots, \mbf f_{p,i}) = \lambda\; \frac{\Vert (\mbf f_{1,i}' \ldots \mbf f_{p,i}')'\Vert_1}{\Vert (\widetilde{\mbf f}_{1,i}' \ldots \widetilde{\mbf f}_{p,i}')'\Vert_1}.
\]
\item [\textit{(iii)}] Group lasso, as defined by \citet{yuan2006model}, where the explanatory variables are grouped before penalising, thus in a VAR context the groups are given by the lags of each variable, thus there are $n$ groups, each of $p$ elements, and we have
\[
\mathcal P^{GL}(\mbf f_{1,i} \ldots, \mbf f_{p,i}) = \lambda \sqrt p \sum_{j=1}^n \Vert (f_{1,ij}\ldots f_{p,ij})'  \Vert_2.
\] 
\end{compactenum}
The penalisation constant $\lambda$, in all three methods, and the maximum VAR lag $p$, are determined by minimising, over a grid of possible values, a BIC-type criterion.

Elastic net and adaptive lasso are particularly useful in a time series context since they are known to stabilise a simple lasso estimator which might suffer on instability due to serial dependence in the data.~To the best of our knowledge,  elastic net  so far has not been considered in the estimation of high-dimensional VARs, while adaptive lasso has been   used by \citet{Kock:Callot:2012} and \citet{nets14}, among others. On the other hand,  group lasso in principle is likely to make the LGCN more sparse than  the other two methods, and has been used, for VAR estimation,  by \citet{nicholson2014hierarchical} and \citet{gelper2015identifying}.~Other possible penalised VAR estimators, which we do not consider here, include the simple lasso and smoothly clipped absolute deviation (SCAD) (see, for example, \citealp{hsu2008subset} and \citealp{abegaz2013sparse}). The consistency of all those  methods is proved in the papers referenced when both $n$ and $T$ tend to infinity, under a variety of technical assumptions which we do not report here.

\subsection{Identification of the LVDN}\label{sec:est3}
Once an estimator of the VAR coefficients is obtained, we can estimate, in a last step, the PCN of the VAR residuals, as defined in \eqref{PCN2}, which in turn can be used for LVDN identification. This  again can be performed by means of several regularisation methods. Here we chose the approach proposed by \citet{Peng:Wang:Zhou:Zhi:2009}, where we refer to for technical details. More precisely, the weights of the PCN are obtained by estimating the regressions in \eqref{PCN3} via  traditional lasso \citep{tibshirani1996regression} in order to ensure sparsity as required by Assumption 4. Alternatively, the PCN of the residuals can be estimated jointly with the LGCN as originally proposed by \citet{rothman2010sparse} for cross-sectional regressions and, in a time series context,   by \citet{abegaz2013sparse}, \citet{gelper2015identifying} or (with  adaptive lasso penalty)  \citet{nets14}. Once we have estimated the PCN of VAR residuals, we can order them according to their centrality in the network. We then estimate the matrix $\mi R_n^T$, needed for identification, as the Choleski factor of the sample covariance matrix of the ordered residuals. Finally, from $\mi R_n^T$ and the estimated VAR operators $\mi F_n^T(L)$, we compute the VMA operator~$\mi D_n^T(L)=(\mi F_n^T(L))^{-1}\mi R_n^T$. The estimated LVDNs weights, denoted as $w_{ij}^{hT}$, readily follow from \eqref{eq:dijh}. 

It has to be noted that, while by estimating a sparse VAR the LGCN  by construction is sparse, this is not the case for the LVDN which,  being derived from the {\it inverse} of a sparse VAR, does not necessarily have to be  sparse. In other words, the assumption of sparsity of VAR coefficients is made for the purpose of dealing with the curse of dimensionality, and does not entail sparsity of the corresponding LVDN. The matrix  $\mi W^{hT}$  with entries $w_{ij}^{hT}$ (the LVDN adjacency matrix) is therefore not necessarily sparse. Nevertheless, since most of its entries are quite  small, considering  a sparse  thresholded version $\mi W^{hT}_{\tau}$  with  threshold $\tau > 0$ is very natural. Here, we chose  the threshold $\tau$   minimising the sum  $\Vert(\mi W^{hT}_{\tau}\mi W^{hT'}_{\tau})^{-1/2}(\mi W^{hT}_{\tau}\mi W^{hT'}_{\tau})(\mi W^{hT}_{\tau}\mi W^{hT'}_{\tau})^{-1/2}-\mi I_n\Vert_2$ of squared errors \citep[see for example][for a similar approach, although in a different context]{FLM13}.


\section{Network analysis of financial volatilities}\label{sec:vol}

The available datasets, in the study of interdependencies of financial institutions, in general, are (large) panels of stock returns. If our interest is in the systematic, i.e.~market-driven, and systemic components of risk, then volatilities are what we need, not returns. Financial crises typically are characterised by unusually high levels of volatility   generated by some major systemic events as the bankruptcy of some major institutions. Analysing interdependencies in volatility  panels  is the first step to a study of  financial contagion \citep[see][for a thorough discussion]{Diebold:Yilmaz:2013}.

Volatilities unfortunately are unobserved, and therefore must be estimated from the panels of returns. Many volatility proxies can be constructed from the series of returns as, for example, the adjusted log-range \citep{parkinson80}   from daily returns,  or log-realised volatilities \citep{ABDL03} based on intra-daily returns. Those proxies are generally treated as observed quantities, and nothing is said about the associated estimation error. On the other hand, volatilities can also be estimated from conditionally heteroschedastic models for financial returns such as multivariate GARCH models \citep[see, for instance, the survey by][]{BLR06}. Those are however parametric models which, due to the curse of dimensionality problem, cannot be handled in the present large-$n$ setting. We therefore follow the approach in \citet{BH15} in adopting a global point of view, with a joint analysis of returns and volatilities in a high-dimensional setting. That analysis is based on a two-step dynamic factor procedure: a first GDFM procedure, applied to the panel of returns, is  extracting  a (double) panel of volatility proxies, which, in a second step, is analysed via a second GDFM. The LVDNs we are interested in are those of the common and idiosyncratic volatility components resulting from this second step.

More precisely, we consider a panel $\bm{r}_n:=\{\bm{r}_{nt}=(r_{1t}\; r_{2t}\ldots r_{nt})'\vert\, t\in\mathbb Z\}$ of $n$ stock returns  (such as the constituents of the S\&P100 index). We assume that $\bm{r}_n$ satisfies Assumptions 1-3, i.e., admits the GDFM  decomposition
\beq
\bm{r}_{nt} = \bm\chi_{nt}+\bm\xi_{nt},\quad t\in\mathbb Z,\label{eq:GDFMret}
\eeq
where   $\bm\chi_{n}$ is driven by $q$ common shocks and $\bm\xi_{n}$ is   idiosyncratic---call them   {\it level-common} and {\it level-idiosyncratic} components, respectively. 

When applied to the real data in Section \ref{sec:emp},  the Hallin and Li\v ska~(2007) criterion very clearly   yields  $q=1$, that is, a unique level-common shock. 
Thus the level-common component $ \bm\chi_{n}$ admits (see  \eqref{eq:ARX}) the autoregressive representation 
\beq
\mathcal A_n(L) \bm\chi_{nt} =: \bm\eta_{nt}=: (\eta_{1t},\ldots , \eta_{nt})^\prime ,\quad t\in\mathbb Z,\label{eq:GDFMret2}
\eeq
where $\mathcal A_n(L)$ is a one-sided square-summable stable block-diagonal autoregressive  filter  with   blocks of size $2\times 2$, and 
$\bm{\eta}_{n}$ is an $n$-dimensional white noise process with a singular    covariance matrix of rank one. 

We then assume that  Assumption 4 holds  for the level-idiosyncratic component $\bm\xi_n$, which thus admits the sparse VAR representation
\beq\label{eq:varidioret}
\mathcal F_n(L)\bm\xi_{nt} = \bm{v}_{nt}=:(v_{1t},\ldots , v_{nt})^\prime ,\quad t\in\mathbb Z
\eeq 
where $\bm{v}_{n}$ is an $n$-dimensional white noise process and $\mathcal F_n(L)$ is a one-sided stable VAR filter with sparse coefficients, the  rows of which have a finite number of  non-zero terms. Estimators of $\bm\eta_n$ and $\bm v_n$ are obtained by applying the methodology described in the previous section.

Turning to volatilities, 
 define  two panels of  volatility proxies, 
\beq\label{eq:volproxy}
\bm \sigma_{nt} := \log(\bm{\eta}_{nt}^2)\quad\text{ and }\quad  \bm\omega_{nt} := \log(\bm{v}_{nt}^2),\quad t\in\mathbb Z , 
\eeq 
with $\log(\bm{\eta}_{nt}^2):=(\log \eta_{1t}^2,\ldots ,\log \eta_{nt}^2)^\prime$ and $\log(\bm{v}_{nt}^2):=(\log v_{1t}^2,\ldots ,\log v_{nt}^2)^\prime$. 
After due centring, we assume that those two panels of volatility proxies in turn satisfy Assumptions~1-3,  and hence admit the GDFM decompositions
\begin{align}
\mathring{\bm\sigma}_{nt}:= \bm\sigma_{nt} -\E[\bm\sigma_{n}] &= \bm\chi_{\sigma,nt}+\bm\xi_{\sigma,nt},\quad t\in\mathbb Z,\label{eq:volmodel_s}\\
\mathring{\bm\omega}_{nt} := \bm\omega_{nt}-\E[\bm\omega_{n}]&=\bm\chi_{\omega,nt}+\bm\xi_{\omega,nt},\quad t\in\mathbb Z,\label{eq:volmodel_o}
\end{align}
where $\bm\chi_{\sigma,n}$ and $\bm\chi_{\omega,n}$ are driven by $q_{\sigma}$ and $q_{\omega}$ common shocks, respectively, and $\bm\xi_{\sigma,n}$ and $\bm\xi_{\omega,n}$ are idiosyncratic---call them the volatility common and idiosyncratic components, respectively. 

Note that traditional factor models for volatilities, derived from factor models of returns, are assuming that $\mathring{\bm\sigma}_{n}$ has no idiosyncratic component and, more importantly,   that $\mathring{\bm\omega}_{n}$ has no common component. Such assumption is quite unlikely to hold,  as there is no reason for market shocks only  affecting  the volatility $\mathring{\bm\sigma}_{n}$ of level-common returns. The empirical results in \citet{BH15} indeed amply confirm that a non-negligible proportion of the variance of $\mathring{\bm\omega}_{n}$ is explained by the same market shocks  also driving~$\mathring{\bm\sigma}_{n}$.

The two volatility common components $\bm\chi_{\sigma,n}$ and $\bm\chi_{\omega,n}$ jointly define a $2n$-dimensional panel made of two blocks. These blocks might be driven by $q_{\sigma\omega}$ shocks, some of which  common in both blocks, and some others   common only in one of them   \citep[see][for a general theory of factor models with block structure]{hallinliska11}. However, when analysing the real data in Section~\ref{sec:emp}, we find that $q_{\sigma\omega}=q_{\sigma}=q_{\omega}=1$, and therefore there is evidence of a unique  common shock, denoted as $\{\varepsilon_t\}$, driving both blocks, and thus unambiguously qualifying as the market volatility shock. The autoregressive representation of these two common components then reads as (see also \eqref{eq:ARX})
\begin{align}\label{eq:AsAo}
\l(\begin{array}{cc}
\mi A_{\sigma,n}(L) & \mi 0_n\\
\mi 0_n & \mi A_{\omega,n}(L)
\end{array}
\r)
\l(
\begin{array}{c}
\bm\chi_{\sigma,n} \\
\bm\chi_{\omega,n}
\end{array}
\r)= \l(\begin{array}{c}
\mi H_{\sigma,n}\\
\mi H_{\omega,n}
\end{array}
\r)\varepsilon_{t},\quad t\in\mathbb Z,\quad \varepsilon_{t}\sim w.n.(0,1),
\end{align}
where $\mi A_{\sigma,n}(L)$ and $\mi A_{\omega,n}(L)$ are one-sided square-summable  block-diagonal stable filters with blocks of size $2\times 2$, and $\mi H_{\sigma,n}$ and $\mi H_{\omega,n}$ are $n$-dimensional column vectors. All parameters in \eqref{eq:AsAo} can be estimated as described in Section \ref{sec:est1}. 

The singular LVDNs for the common components of volatilities thus can be built from  the VMA filters (see  \eqref{eq:LVDNcom4})
\begin{equation}\label{eq:nets1}
\mi B_{\sigma,n}(L) :=(\mi A_{\sigma,n}(L))^{-1}\mi H_{\sigma,n} K_{\sigma}  \quad\text{and}\quad 
\mi B_{\omega,n}(L) :=(\mi A_{\omega,n}(L))^{-1}\mi H_{\omega,n} K_{\omega},
\end{equation}
where $ K_{\sigma}$ and $ K_{\omega}$ in this case are just scalars needed to identify the scale and sign of the market shocks. For a given horizon $h$, the LVDN  weights defined in \eqref{eq:dijh}  provide  the percentages of $h$-step ahead forecast error variance of series $i$ accounted for by the common market shock.

For the two volatility idiosyncratic components $\bm\xi_{\sigma,n}$ and $\bm\xi_{\omega,n}$, we assume that Assumption~4 holds, yielding the sparse VAR representations
\begin{align}
\mi F_{\sigma,n}(L)\bm\xi_{\sigma,n}&=\bm\nu_{\sigma,nt},\quad t\in\mathbb Z,\quad \bm\nu_{\sigma,nt}\sim w.n.(\mbf 0,\mi C_{\sigma,n}^{-1}),\label{eq:Fs}\\
\mi F_{\omega,n}(L)\bm\xi_{\omega,n}&=\bm\nu_{\omega,nt},\quad t\in\mathbb Z,\quad \bm\nu_{\omega,nt}\sim w.n.(\mbf 0,\mi C_{\omega,n}^{-1}),\label{eq:Fo}
\end{align}
where $\mi F_{\sigma,n}(L)$ and $\mi F_{\omega,n}(L)$  are one-sided stable filters with sparse coefficients, the  rows of which have a finite number of non-zero terms. All parameters in \eqref{eq:Fs} and \eqref{eq:Fo} can be estimated as described in Section \ref{sec:est2}. 

Inverting those autoregressive representations, we obtain the VMA filters (see also \eqref{eq:LVDNidio00})
\begin{equation}\label{eq:nets2}
\mi D_{\sigma,n}(L) :=(\mi F_{\sigma,n}(L))^{-1}\mi R_{\sigma,n} \quad\text{and}\quad  
\mi D_{\omega,n}(L) :=(\mi F_{\omega,n}(L))^{-1}\mi R_{\omega,n}.
\end{equation}
Here,   $\mi R_{\sigma,n}$ and $\mi R_{\omega,n}$ are $n\times n$ invertible matrices such that $\mi R_{\sigma,n}^{-1}\bm\nu_{\sigma,n}$ and 
$\mi R_{\omega,n}^{-1}\bm\nu_{\omega,n}$ are orthonormal. As explained in Section~\ref{sec:gdfm}, we choose those matrices to be  
the Choleski factors of the covariances of the VAR shocks $\bm\nu_{\sigma,n}$ and $\bm\nu_{\omega,n}$, ordered according to their centrality in the PCN induced by $\mi C_{\sigma,n}$ and $\mi C_{\omega,n}$ (see also \eqref{PCN1}-\eqref{PCN3}). From \eqref{eq:nets2}, and for a given horizon~$h$, the LVDN weights computed from  \eqref{eq:dijh}  then provide the   shares of  $h$-step ahead forecast error variance for the idiosyncratic volatility of series $i$ accounted for by innovations in the idiosyncratic volatility of series $j$. As already explained in Section \ref{sec:est3}, most of those weights are quite small, and naturally can be thresholded, yielding a sparse network. 



\section{The network of S\&P100 volatilities}\label{sec:emp}

In this section, we consider the panel of stocks used in the construction of  the S\&P100 index and, based on the daily adjusted closing prices $\{p_{it}\vert\ i=1,\ldots,n,  \  t=1,\ldots, T\}$, we compute the panel of percentage daily log-returns 
\beq
r_{it} :=100\log \l({p_{it} / p_{it-1}}\r),\quad i=1,\ldots ,n,\quad t=1,\ldots, T,\nn
\eeq
which are our observed  ``levels'' or ``returns''. The observation period is $T=3457$ days, from 3rd January 2000 to  30th September 2013 and,  since not all 100 constituents of the index were traded during the  observation period, we end up with a panel of $n=90$ time series. A detailed list of the series considered is provided in the Web-based supporting materials. All results in this section are reported for the whole period 2000-2013, and for the period 2007-2008 corresponding to the Great Financial Crisis. Networks are represented by heat-maps showing the entries of the corresponding adjacency matrices. In all those plots we highlight the Energy and Financial sectors which correspond to the index values 22-33 and 34-46, respectively. Moreover, all LVDNs considered  report shocks' effects over $h=20$ periods, corresponding to a one-month horizon (20 days plus the contemporaneous effects). Results for $h=5,10$ are reported in the Web-based supporting materials.

As already explained,   the \citet{hallinliska07} criterion yields $q^T=1$, that is, the level-common component is driven by a one-dimensional common shock. By estimating \eqref{eq:GDFMret} and \eqref{eq:GDFMret2}, we recover the estimated level-common component $\bm\chi_n^T$, the rank-one shocks $\bm{\eta}_{n}^T$, and the level-idiosyncratic component $\bm\xi_n^T$. The contribution of the level-common component to the total variation of returns, computed as the ratio between the sum of the (empirical) variances of the estimated level-common components ${\bm \chi}_n^T$ to the sum of the (empirical) variances of the observed returns, is 0.36. Turning to the level-idiosyncratic component $\bm\xi_n^T$, we recover an~$n$-dimensional innovation vector~$\bm{v}_n^T$ by fitting a sparse VAR.

The panels $\bm\sigma_{n}^T$ and  $\bm\omega_{n}^T$ of level-common and level-idiosyncratic volatilities    are constructed, as explained  in \eqref{eq:volproxy}, from  the $\bm{\eta}_{n}^T$'s and $\bm{v}_n^T$'s;   the centred panels $\mathring{\bm\sigma}_{n}^T$ and $\mathring{\bm\sigma}_{n}^T$ are obtained by subtracting the sample means. Applying the \citet{hallinliska07} criterion again, we obtain~$q_{\sigma}^T=q_{\omega}^T=1$ and, for the global panel,  $q_{\sigma\omega}^T=1$. This implies a unique market shock, common to the two sub-panels. We then compute the estimators $\bm\chi_{\sigma,n}^T$ and  $\bm\xi_{\sigma,n}^T$ of 
the two level-common volatility components of the GDFM decomposition~\eqref{eq:volmodel_s}, and the estimators $\bm\chi_{\omega,n}^T$ and~$\bm\xi_{\omega,n}^T$  of the two level-idiosyncratic volatility components of the GDFM decomposition~\eqref{eq:volmodel_o}. The overall contribution of the common components (driven by the unique market shock, hence non-diversifiable) to the total variances of $\bm\sigma_{n}^T$ and $\bm\omega_{n}^T$ are $0.60$ and $0.17$ respectively.

\subsection{The LVDN of volatility common components}
We now turn to the LVDNs of the two common components $\bm\chi_{\sigma,n}^T$ and $\bm\chi_{\omega,n}^T$, given by \eqref{eq:nets1}. Since both panels are driven  by the same unique common shock, the networks are identified once we impose a sign and a scale on the shocks. The sign is set in such a way that the sample correlation between the estimated market shock $\{\varepsilon ^T_t\}$  and the cross-sectional average of all common components is positive. The scale is set in such a way that the following results represent the effect of one-standard-deviation market volatility shock, that is, the sequence of MA loading coefficients is divided by the standard error of the market shock. We then  obtain the estimated filters $\mi B_{\sigma,n}^T(L)$ and $\mi B_{\omega,n}^T(L)$. 

Since LVDNs in this case are singular, we do not show a graph, but  rather report, in Table~\ref{tab:LVDN_common}, for each panel~$\bm\chi_{\sigma,n}^T$ and $\bm\chi_{\omega,n}^T$,  the percentages of sectoral $20$-step ahead forecast error variance accounted for by the unique market shock. More precisely, for the ten sectors considered, the figures in Table~\ref{tab:LVDN_common}  are the ratios 
\begin{equation}
w_{\chi_\sigma,j}^T :=  100\l(\frac{\sum_{i=1}^{n_j}\sum_{k=0}^{20} (b_{\sigma,k,i}^T)^2 }{\sum_{i=1}^n\sum_{k=0}^{20} (b_{\sigma,k,i}^T)^2 }\r)  \text{ and }w_{\chi_\omega,j}^T :=  100\l(\frac{\sum_{i=1}^{n_j}\sum_{k=0}^{20} (b_{\omega,k,i}^T)^2 }{\sum_{i=1}^n\sum_{k=0}^{20} (b_{\omega,k,i}^T)^2 }\r),\quad j = 1,\ldots ,10,\nn
\end{equation}
where $b_{\sigma,k,i}^T$ and $b_{\omega,k,i}^T$ are the $i$-th entries of $\mi B_{\sigma,nk}^T$ and $\mi B_{\omega,nk}^T$, respectively, and $n_j$ is the number of stocks in sector $j$.
As  the single shock obviously explains all the variance of the common component, we normalise the figures in each column so that their sum equals one hundred. 

In both periods considered, the common component of level-common volatility is affected uniformly across all sectors by a market shock (the $w_{\chi_\sigma,j}^{T}$ columns  in Table \ref{tab:LVDN_common}), while the common components of level-idiosyncratic volatilities  exhibit some interesting inter-sectoral differences (the $w_{\chi_\omega,j}^{T}$ columns  in Table \ref{tab:LVDN_common}). In particular, when looking at the $w_{\chi_\omega,j}^{T}$ column,  the Energy and Financial sectors are  the most affected ones. The impact of a market shock during the crisis  is even heavier on those two sectors. We refer to \citet{BH15} for further results on the volatility of common components.
\begin{table}[t!]
\begin{center}
\footnotesize{
\begin{tabular}{lrr|rr}
\hline
\hline
	& \multicolumn{2}{c}{2000-2013}& \multicolumn{2}{|c}{2007-2008}\\
Sector&  $w_{\chi_\sigma,j}^{T}$ & $w_{\chi_\omega,j}^{T}$&$w_{\chi_\sigma,j}^{T}$ & $w_{\chi_\omega,j}^{T}$\\
\hline
Consumer Discretionary&10.10&7.63&9.91&7.73\\
Consumer Staples&10.38&10.70&9.83&10.45\\
Energy&9.86	&13.35&9.83&27.04\\
Financial&10.12&13.66&10.61&18.19\\
Health Care&9.92&8.84&9.77&6.24\\
Industrials&9.41&7.59&9.59&6.34\\
Information Technology&10.01&10.00&10.11&3.76\\
Materials&9.92&6.77&10.05&9.53\\
Telecommunication Services&10.33&9.80&10.29&5.79\\
Utilities&9.96&11.67&10.01&4.93\\
\hline
Total & 100&100&100&100\\
\hline
\hline
\end{tabular}
}\vspace{-2mm}
\end{center}
\caption{\textnormal{Percentages of 20-step ahead forecast error variances due to the market shock.}\vspace{-3mm}}\label{tab:LVDN_common}
\end{table}

\subsection{The LVDN of the volatility idiosyncratic components}
Turning to the idiosyncratic volatilities $\bm\xi_{\omega,n}^T$ and $\bm\xi_{\sigma,n}^T$, it appears that $\bm\xi_{\sigma,n}^T$  is essentially uncorrelated, both   serially and cross-sectionally. Therefore, we focus only on the idiosyncratic volatility $\bm\xi_{\omega,n}^T$ of  level-idiosyncratic components. Based on a BIC criterion, we estimate a sparse VAR(5) for $\bm\xi_{\omega,n}$.
The weight associated with the $(i,j)$ edge of the LGCN of $\bm\xi_{\omega,nt}^T$ then  is   the $(i,j)$ entry of $\mi F_{\omega,n}^T(1)=\sum_{k=0}^5 \mi F_{\omega,n k}^T$. 

\begin{figure}[t!]
	\begin{center}
	\begin{tabular}{cc}	
	\vspace{-6mm}\\
	{\includegraphics[width=0.3\textwidth]{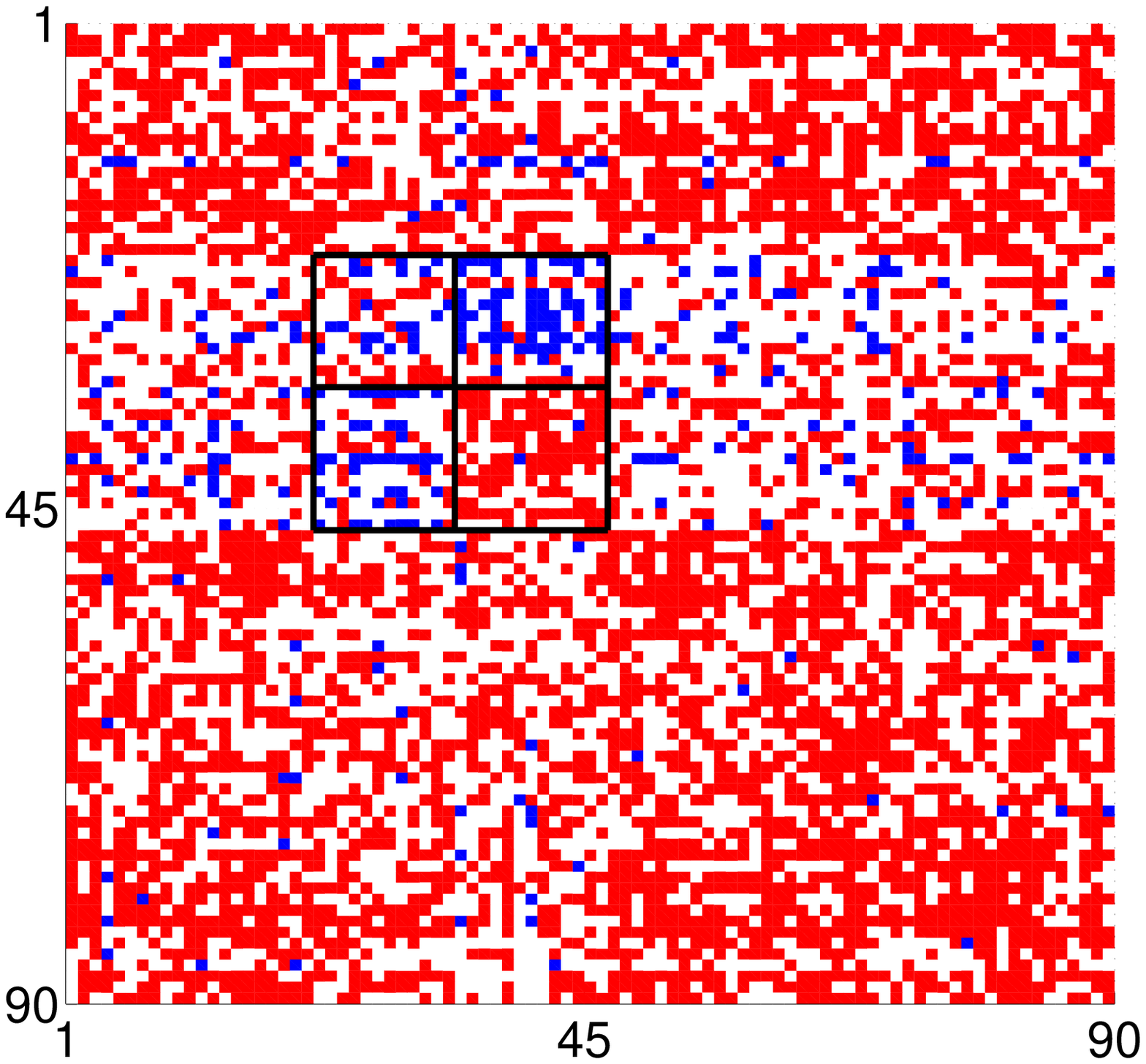}}	&	{\includegraphics[width=0.3\textwidth]{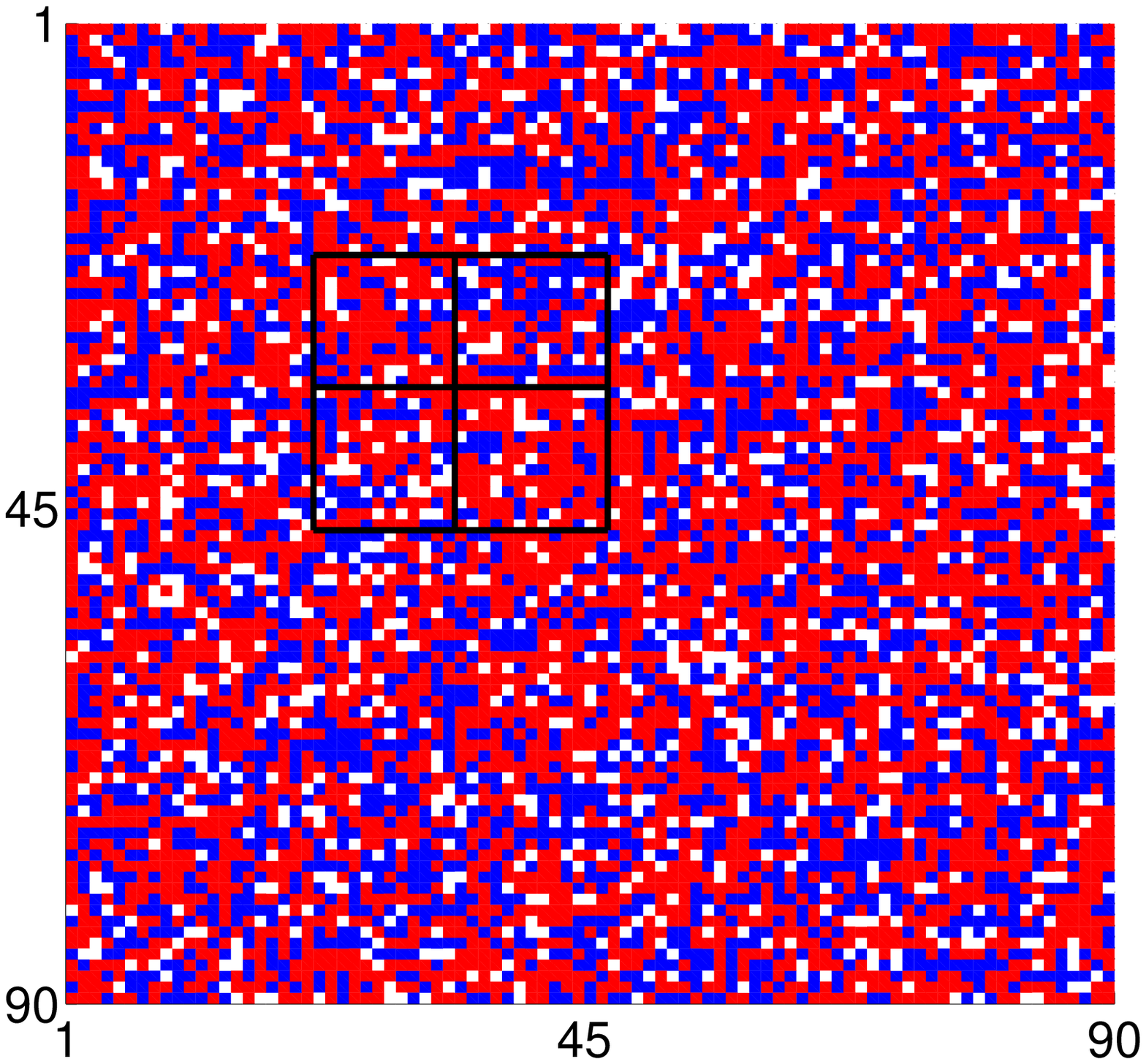}}	\\
	\vspace{-12mm}\\
			\footnotesize{2000-2013} &\footnotesize{2007-2008}\\
	\end{tabular}
	\vspace{-2mm}
	\end{center}
	\caption{\textnormal{LGCN for $\bm\xi_{\omega,n}^T$, negative weights in blue, the positive ones in red.}\vspace{-3mm}} \label{fig:LGCN_oo}
\end{figure}

We define the  {\it edge density} of a network with set of edges $\cal E$ as the proportion $\# ({\mathcal E})/(n^2-n)$ of couples $(i,j)$ in $\cal E$. When considering the elastic net estimation approach, we obtain, for the period 2000-2013,  
a LGCN with edge density~$53\%$ whereas, for the period 2007-2008, that density is close to $86\%$. The corresponding LGCNs are shown on Figure \ref{fig:LGCN_oo}. Note that, as expected, the LGCNs are much  sparser when resulting from group lasso estimation, with densities $14\%$ for the period 2000-2013, and $37\%$ for the period 2007-2008. Nevertheless, since the subsequent results on LVDNs turn out to be qualitatively similar regardless of the way we estimate the VAR, we only show here   the elastic net results, deferring      group and adaptive lasso results  to the Web-based supporting materials. 

\begin{figure}[t!]
	\begin{center}
	\begin{tabular}{cc}	
	\vspace{-6mm}\\
	{\includegraphics[width=0.3\textwidth]{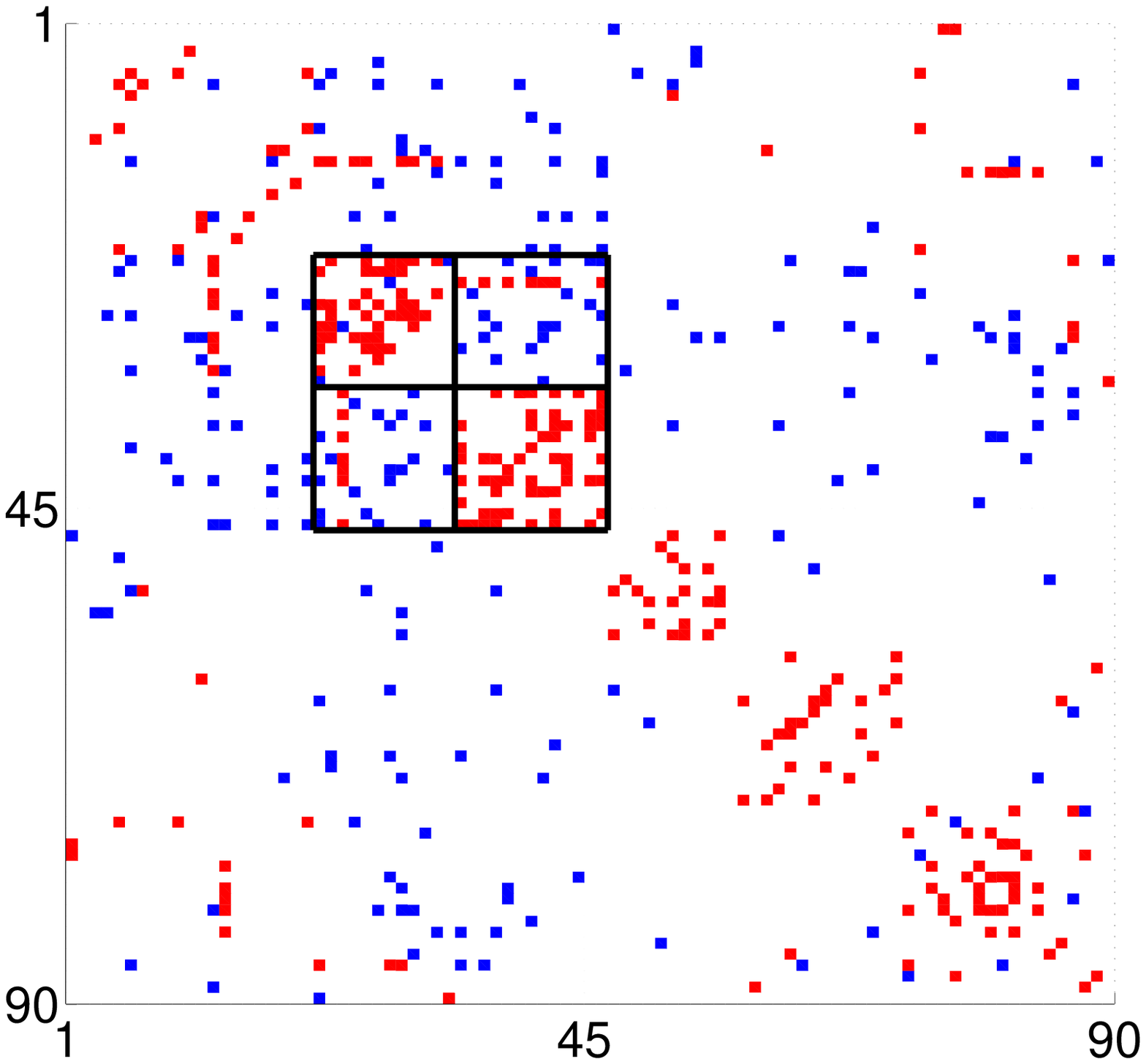}}	&	{\includegraphics[width=0.3\textwidth]{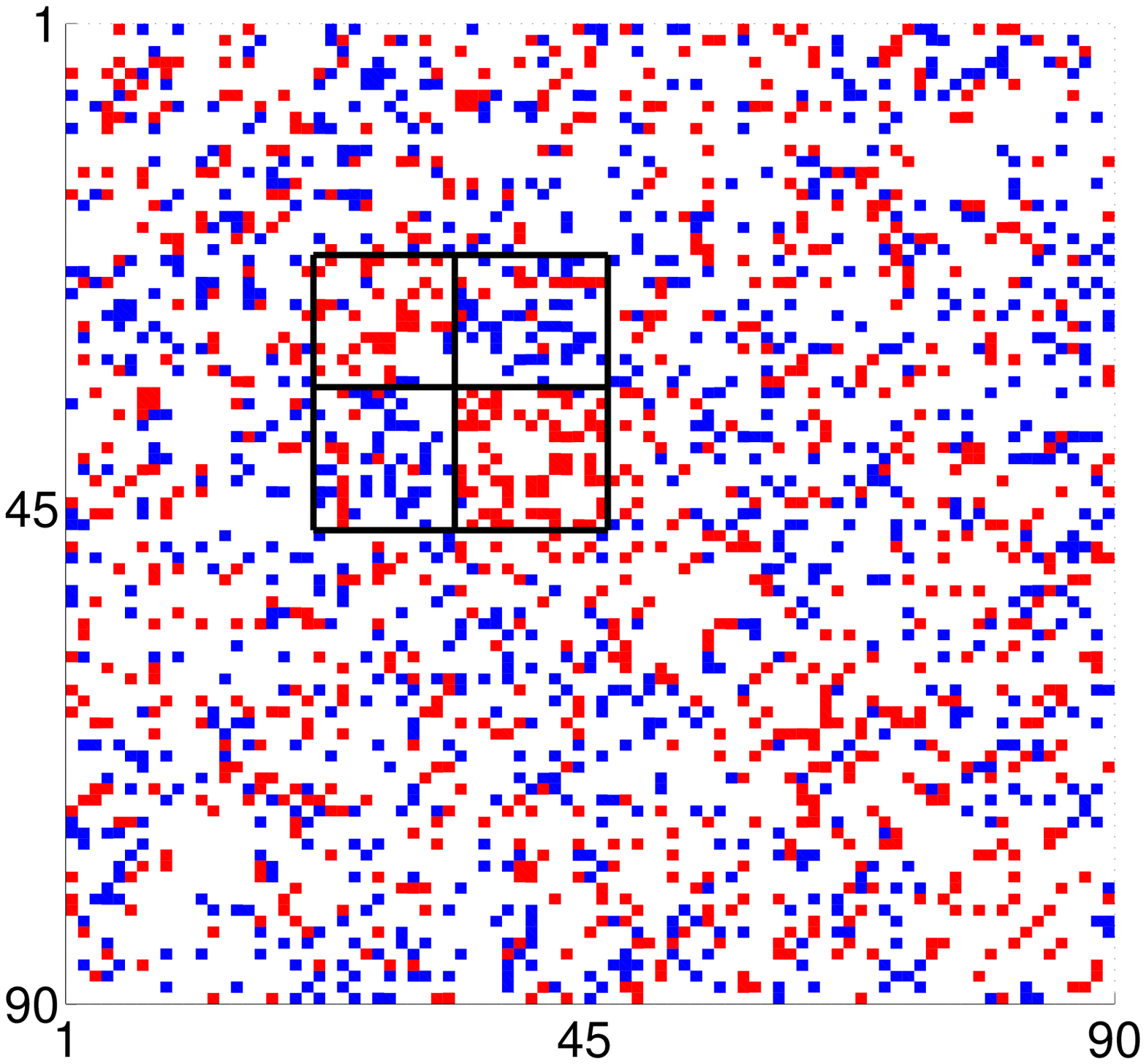}}	\\
	\vspace{-12mm}\\
	\footnotesize{2000-2013} &\footnotesize{2007-2008}\\
	\end{tabular}
	\vspace{-2mm}
	\end{center}
	\caption{\textnormal{PCN for $\bm \nu_{\omega,n}^T$, negative weights in blue, the positive ones in red.}} \label{fig:PCN_oo}
\end{figure}

As explained in the previous sections, identification of the LVDN requires some choices. The Choleski ordering is appealing, since it is data-driven, but requires an ordering of the cross-sectional items (the stocks). Here, we use an identification method based on the partial correlation of the VAR residuals $\bm{v}_{\omega,n}^T$.~As explained in Section 2, we ordered the stocks according to the concept of eigenvector centrality for undirected networks \citep[see][]{bonacich1987power} in the estimated PCN. The resulting network is shown in Figure \ref{fig:PCN_oo} for the two periods under study. The densities of those networks are~6\% for the period 2000-2013 and 24\% for the period 2007-2008. The ten most central stocks are reported in Table \ref{tab:evc}. For results using other identification methods we refer to the Web-based supporting material. 

Summing up, from the estimation of the VAR model and the analysis of its residuals we see that (i) the Great Financial Crisis has considerably blown up  the dynamic interdependencies between stocks, and (ii) the Energy and Financial stocks appear as the most interconnected ones, with more intra-sectoral dependencies rather than inter-sectoral. 

\begin{table}[t!]
\begin{center}
\footnotesize{
\begin{tabular}{ l | l }
\hline
\hline
{2000-2013}&{2007-2008}\\
\hline
JPM	JP Morgan Chase \& Co.	&	BAC	Bank of America Corp.	\\
C	Citigroup Inc.	&	USB	US Bancorp	\\
BAC	Bank of America Corp.	&	JPM	JP Morgan Chase \& Co.	\\
APA	Apache Corp.	&	MS	Morgan Stanley	\\
WFC	Wells Fargo	&	WFC	Wells Fargo	\\
COP	Conoco Phillips	&	DVN	Devon Energy	\\
OXY	Occidental Petroleum Corp.	&	GS	Goldman Sachs	\\
DVN	Devon Energy	&	AXP	American Express Inc.	\\
SLB	Schlumberger	&	COF	Capital One Financial Corp.	\\
MS	Morgan Stanley	&	UNH	United Health Group Inc.	\\
\hline
\hline
\end{tabular}
\vspace{-2mm}
}
\end{center}
\caption{\textnormal{Eigenvector centrality in the PCN for $\bm \nu_{\omega,n}^T$.}}\label{tab:evc}
\end{table}

\begin{table}[t!]
\begin{center}
\footnotesize{
\begin{tabular}{c c c c c c |c}
\hline
\hline
{\footnotesize percentiles} & 50$^{\mbox{\tiny {th}}}$&	90$^{\mbox{\tiny {th}}}$&	95$^{\mbox{\tiny {th}}}$&	97.5$^{\mbox{\tiny {th}}}$&	99$^{\mbox{\tiny {th}}}$ & {\footnotesize max}\\
\hline
2000-2013& 0.02&	0.13&	0.20&	0.29&	0.48 & 4.29\\
 \hline
 2007-2008& 0.17&	0.71&	1.00&	1.28&	1.76 & 4.53\\
\hline
\hline
\end{tabular}
\vspace{-2mm}
}
\end{center}
\caption{\textnormal{Selected percentiles of  $\bm\xi_{\omega,n}^T$  LVDN weights.}\vspace{-3mm}}\label{tab:quant}
\end{table}

The LVDN for $\bm\xi_{\omega,n}^T$ now can be computed on the basis of the ordering (partially) shown in Table~\ref{tab:evc}. That  identification defines (see~\eqref{eq:nets2}) the matrix $\mi R_{\omega,n}^T$ as the Choleski factor of the sample covariance matrix of the ordered shocks. The weight of    edge $(i,j)$ of the $\bm\xi_{\omega,n}^T$ LVDN is  
\begin{align}
w_{\xi_\omega,ij}^T =  100\l(\frac{\sum_{k=0}^{20} (d_{\omega,k,ij}^T)^2 }{\sum_{j=1}^n\sum_{k=0}^{20} (d_{\omega,k,ij}^T)^2 }\r),\quad i,j = 1,\ldots ,n,\nn
\end{align}
where $d_{\omega,k,ij}^T$ is the $(i,j)$ entry of $\mi D_{\omega,nk}^T$ such that $\mi D_{\omega,n}^T(L)=\sum_{k=0}^{20}\mi D^T_{\omega,nk}L^k$ as defined in~\eqref{eq:nets2}. The resulting network is directed and weighted, but it is not sparse, meaning that in principle all its weights can be different from zero. Still, only a small proportion of edges has weights larger than one, corresponding to a proportion of variance larger than 1\%. 
In particular, out of the total number  ($n(n-1)=8010$) of possible edges, only 0.4\% have weights larger than~1\%  over the period 2000-2013, while this number increases considerably, up to 5\%, during the period 2007-2008.
Selected percentiles   are given in Table \ref{tab:quant}.

\begin{figure}[t!]
	\begin{center}
	\begin{tabular}{cc}	
	\vspace{-6mm}\\
	{\includegraphics[width=0.3\textwidth]{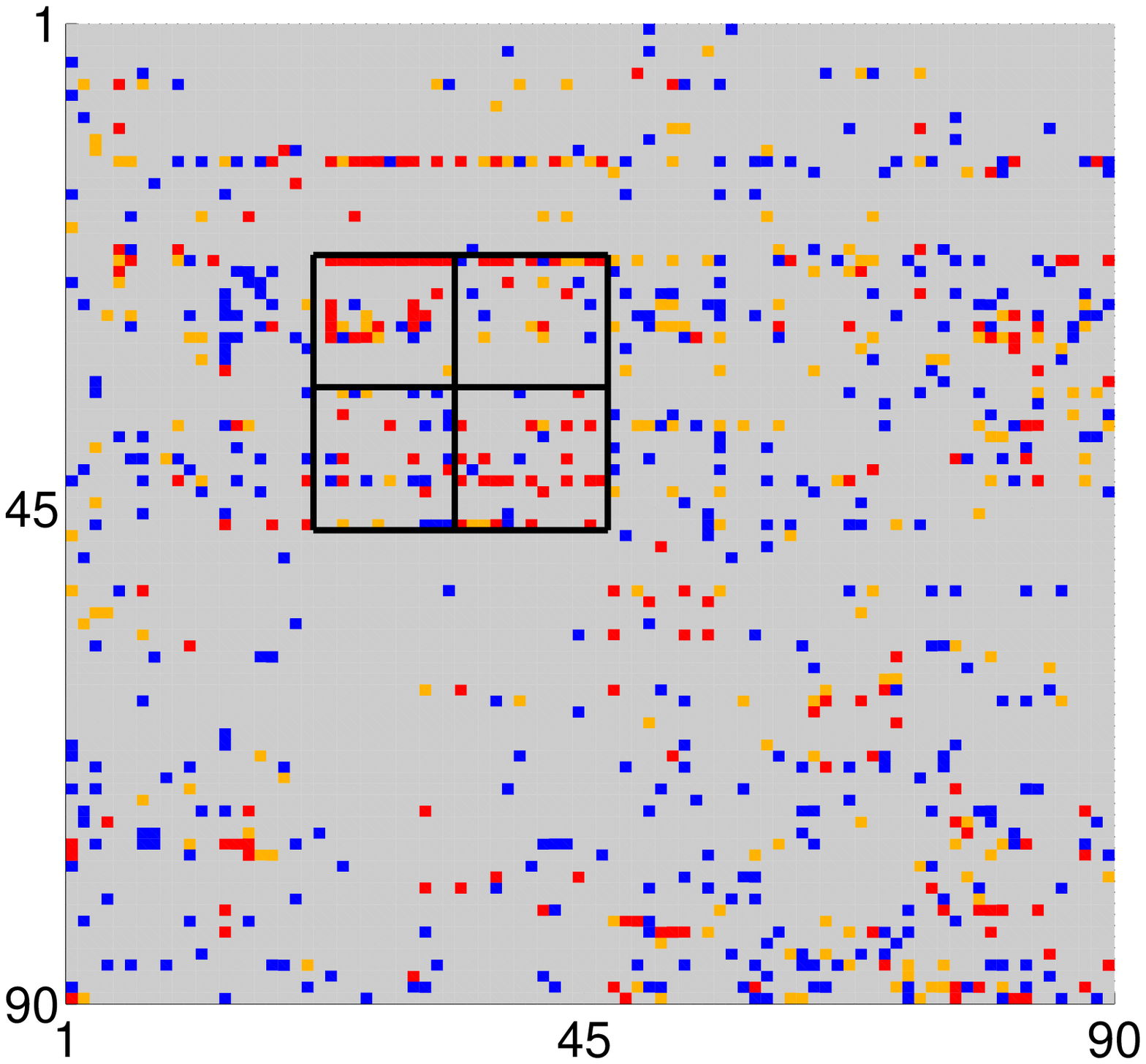}}	&	
	{\includegraphics[width=0.3\textwidth]{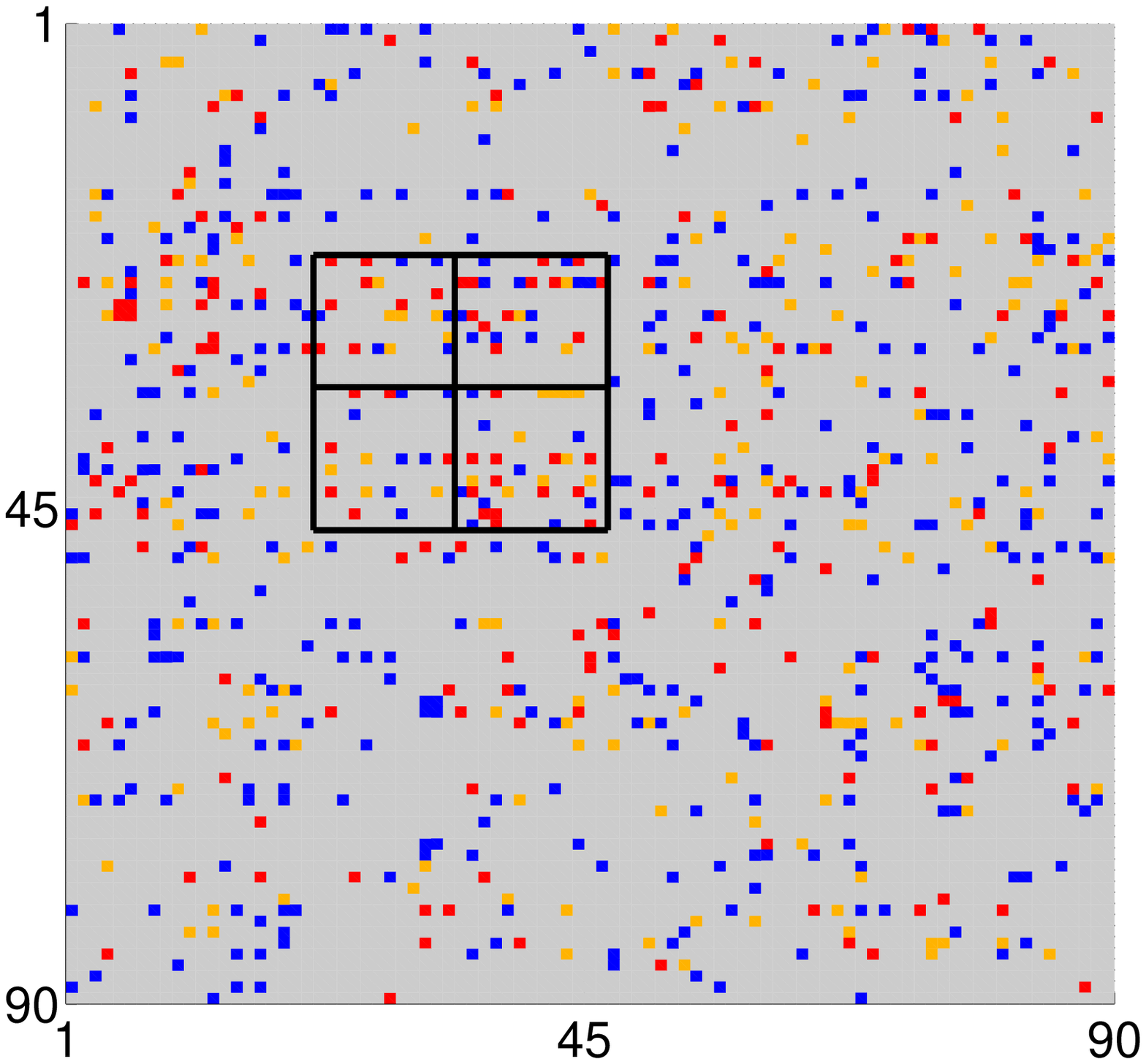}}
\\	

		\vspace{-12mm}\\
			\footnotesize{2000-2013} &\footnotesize{2007-2008}\\
	\end{tabular}
		\vspace{-2mm}
	\end{center}
		\caption{\textnormal{LVDN for $\bm\xi_{\omega,n}^T$, weights below the~95$^{\mbox{\tiny {th}}}$ percentile in grey, between the~95$^{\mbox{\tiny {th}}}$ and~97.5$^{\mbox{\tiny {th}}}$ percentiles in  blue, between the~97.5$^{\mbox{\tiny {th}}}$  and~99$^{\mbox{\tiny {th}}}$ percentiles in yellow, and above the~99$^{\mbox{\tiny {th}}}$ percentile in red.}} \label{fig:VDN_oo}
\end{figure}

\begin{figure}[t!]
	\begin{center}
	\begin{tabular}{cc}	
	\vspace{-6mm}\\
	{\includegraphics[width=0.3\textwidth]{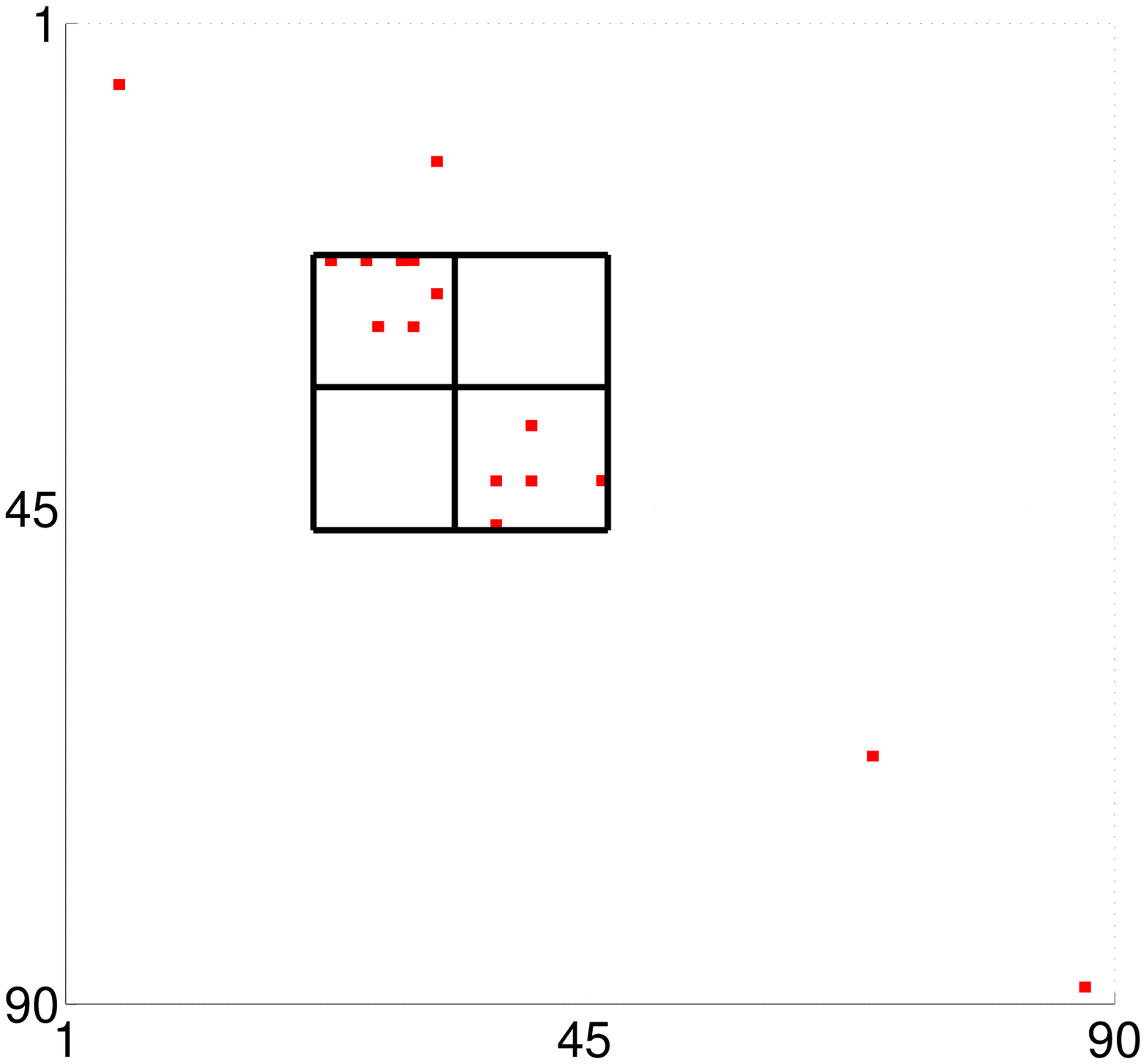}}	&
	{\includegraphics[width=0.3\textwidth]{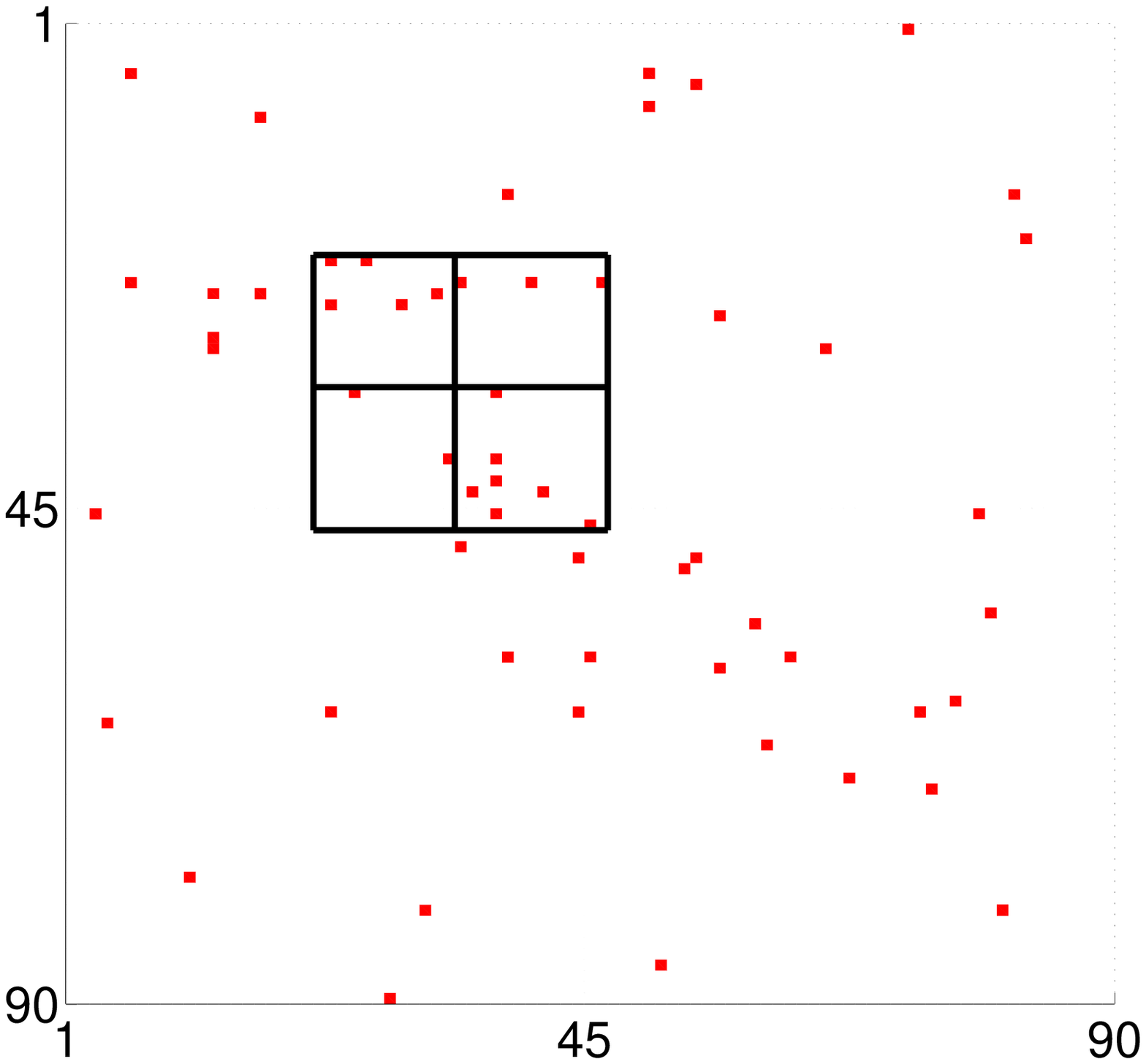}}\\
	
	\vspace{-12mm}\\
	\footnotesize{2000-2013} &\footnotesize{2007-2008}\\
	\end{tabular}
		\vspace{-2mm}
	\end{center}
		\caption{\textnormal{Thresholded LVDN for $\bm\xi_{\omega,n}^T$, non-zero weights in red.}} \label{fig:VDN_oo_thresh}
\end{figure}

Figure \ref{fig:VDN_oo}  shows the $\bm\xi_{\omega,n}^T$ LVDN weights. Inspection reveals that  LVDNs, although  not sparse, have many entries close to zero. A thresholded version,    as described in Section \ref{sec:est},  is reported in the top row of Figure~\ref{fig:VDN_oo_thresh}. The resulting plots are highly sparse and indeed the optimal value of the thresholding parameter $\tau$ is found to be 1.61 for the period 2000-2013 and 1.9 for the period 2007-2008. Plots for other values of $\tau$ are provided  in the Web-based supporting materials. Given the few remaining links, this is conveniently visualised, and  the corresponding networks are shown in Figure~\ref{fig:VDN_oo_thresh}. Note that Financial (yellow nodes) and Energy (blue nodes) stocks are the most interconnected ones. When considering the whole sample, there are almost no inter-sectoral links; however, during the Great Financial Crisis, the degree of interconnectedness of the Financial sector with other sectors quite  dramatically increases.

Those  findings can be quantified by computing  from-  and to-degrees, as defined in~\eqref{eq:indeg}. As explained in Section \ref{sec:gdfm}, the from-degree measures the exposure of a given firm to shocks coming from all other firms, while the to-degree measures the effect of a shock to a given firm on all other ones. In Table~\ref{tab:nets_sec}, we show sectoral averages of from- and to-degrees when computed both for the non-sparse and thresholded LVDNs. 
We also report sectoral total connectedness and overall total connectedness as defined in \eqref{eq:totdeg}. Finally, an overall measure of how systemic is an institution is given by its centrality in the network. Table~\ref{tab:evec2} shows,  for the two periods considered, the rankings of firms according to a measure of eigenvector centrality adapted to weighted directed networks \citep[see][]{bonacich2001eigenvector2}. For the period 2000-2013, only very few stocks are connected and qualify as central.  

All results amply demonstrate that the Great Financial Crisis quite spectacularly  tightened up the links between firms.  In particular, when considering the thresholded   LVDN version, the shocks to Financial stocks still account for almost 7\% of the total idiosyncratic variance during the crisis, while shocks from other sectors explain about 5\% of the Financial intra-sectoral variability (see the last two columns of Table \ref{tab:nets_sec}). The Financial sector appears to be  the most central  in the network,  the most vulnerable to shocks coming from all other sectors, and  the most systemic in the sense that a shock to the Financial stocks  is most likely to strongly affect the whole panel. 

\begin{table}[t!]
\begin{center}
\footnotesize{
\begin{tabular}{l rr | rr | rr | rr}
\hline
\hline
& \multicolumn{4}{c}{\footnotesize non-thresholded}&\multicolumn{4}{|c}{\footnotesize thresholded}\\
	& \multicolumn{2}{c}{2000-2013}& \multicolumn{2}{|c}{2007-2008}& \multicolumn{2}{|c}{2000-2013}& \multicolumn{2}{|c}{2007-2008}\\
Sector&from&to&from&to&from&to&from&to  \\
\hline
Consumer Discretionary	&	4.32	&	2.37	&	26.31	&	26.41	&	0.24	&	0.24	&	1.96	&	3.29	\\
Consumer Staples	&	3.98	&	4.65	&	27.47	&	22.65	&	0.00	&	0.44	&	2.80	&	1.36	\\
Energy	&	5.52	&	7.92	&	21.91	&	33.72	&	1.90	&	1.54	&	4.01	&	6.10	\\
Financial	&	4.74	&	6.22	&	24.42	&	35.56	&	0.77	&	0.77	&	5.23	&	6.69	\\
Health Care	&	5.00	&	2.51	&	28.06	&	22.36	&	0.00	&	0.00	&	4.27	&	1.83	\\
Industrials	&	4.43	&	3.21	&	27.26	&	25.81	&	0.21	&	0.21	&	2.43	&	3.53	\\
Information Technology	&	5.03	&	4.89	&	29.90	&	19.98	&	0.00	&	0.00	&	3.35	&	1.58	\\
Materials	&	3.24	&	4.62	&	26.86	&	27.01	&	0.00	&	0.00	&	2.99	&	0.81	\\
Telecommunications Services	&	6.50	&	7.26	&	27.44	&	16.52	&	1.91	&	1.91	&	0.00	&	0.00	\\
Utilities	&	5.15	&	8.74	&	29.49	&	21.54	&	0.00	&	0.00	&	4.94	&	2.38	\\
\hline
Total degree&4.73	&&	26.54&&		0.47	&&		3.40			\\
\hline
\hline
\end{tabular}
\vspace{-2mm}
}
\end{center}
\caption{\textnormal{From- and To-degree sectoral averages in the  $\bm\xi_{\omega,n}^T$ LVDN.}}\label{tab:nets_sec}
\end{table}

\begin{table}[t!]
\begin{center}
\footnotesize{
\begin{tabular}{ l | l }
\hline
\hline
{2000-2013}&{2007-2008}\\
\hline
BAC	Bank of America Corp.				&	BAC	Bank of America Corp.					\\
JPM	JP Morgan Chase \& Co.				&	AIG	American International Group Inc.					\\
C	Citigroup Inc.				&	COF	Capital One Financial Corp.					\\
WFC 	Wells Fargo				&	USB	US Bancorp					\\
-			&	C	Citigroup Inc.					\\
-			&	WFC	 Wells Fargo					\\
-			&	CVX	Chevron					\\
-			&	LOW	 Lowes					\\
-			&	BA	Boeing Co.					\\
-			&	IBM	International Business Machines					\\
\hline
\hline
\end{tabular}
\vspace{-2mm}
}
\end{center}
\caption{\textnormal{Eigenvector centrality in the  $\bm\xi_{\omega,n}^T$ thresholded LVDN.}\vspace{-3mm}}\label{tab:evec2}
\end{table}

A few more comments  are in order in relation with the results in the Appendix.  First, when comparing the results for $h=20$ with those obtained at shorter horizons ($h=5,\ 10$), we see that connectivity increases considerably with the forecast horizon, indicating that the effect of a shock keeps on propagating for a long time. This result is consistent with the fact that  volatilities in financial data, although  stationary,  tend  to have long memory \citep[see][for a detailed analysis of this aspect]{BH15}. Second,   group lasso and elastic net   VAR estimation yield qualitatively similar results. However, despite of sparser LGCNs, group lasso yields  estimated LVDNs that are more tightly connected than those resulting from elastic net. This is particularly true during the crisis period when   Financial and Energy stocks are the most central ones, but also the Health Care sector seems to have a decisive role.

To conclude, we provide some empirical justification of the ``factor plus sparse VAR'' approach adopted in this paper by comparing the conditional dependencies in the volatility panel~$\bm\omega_n^T$ with those of its idiosyncratic component $\bm\xi_{\omega;n}^T$. To do this, we consider {\it partial spectral coherence} ($PSC$), which is the analogous of partial correlation, but in the spectral domain, and  is strictly related with the coefficients of a VAR representation \citep[see][for a definition]{Davis:Zang:Zheng:2012}. 

 In line with the long-run spirit of the LVDN definition, and since volatilities have strong persistence, we first consider the $PSC$s at frequency $\theta=0$, thus looking at long-run conditional dependencies. Selected percentiles of the distributions of the absolute value of the PSC entries  for $\bm\omega_n^T$ and $\bm\xi_{\omega;n}^T$ and the distribution of the absolute value of their differences are shown in Table~\ref{tab:psc}. Both $PSC$s have many small (in absolute value)  entries, which is  consistent with our sparsity assumptions. 

\begin{figure}[t!]
	\begin{center}
	\begin{tabular}{ccc}	
	\vspace{-10mm}\\	
	{\includegraphics[width=0.3\textwidth]{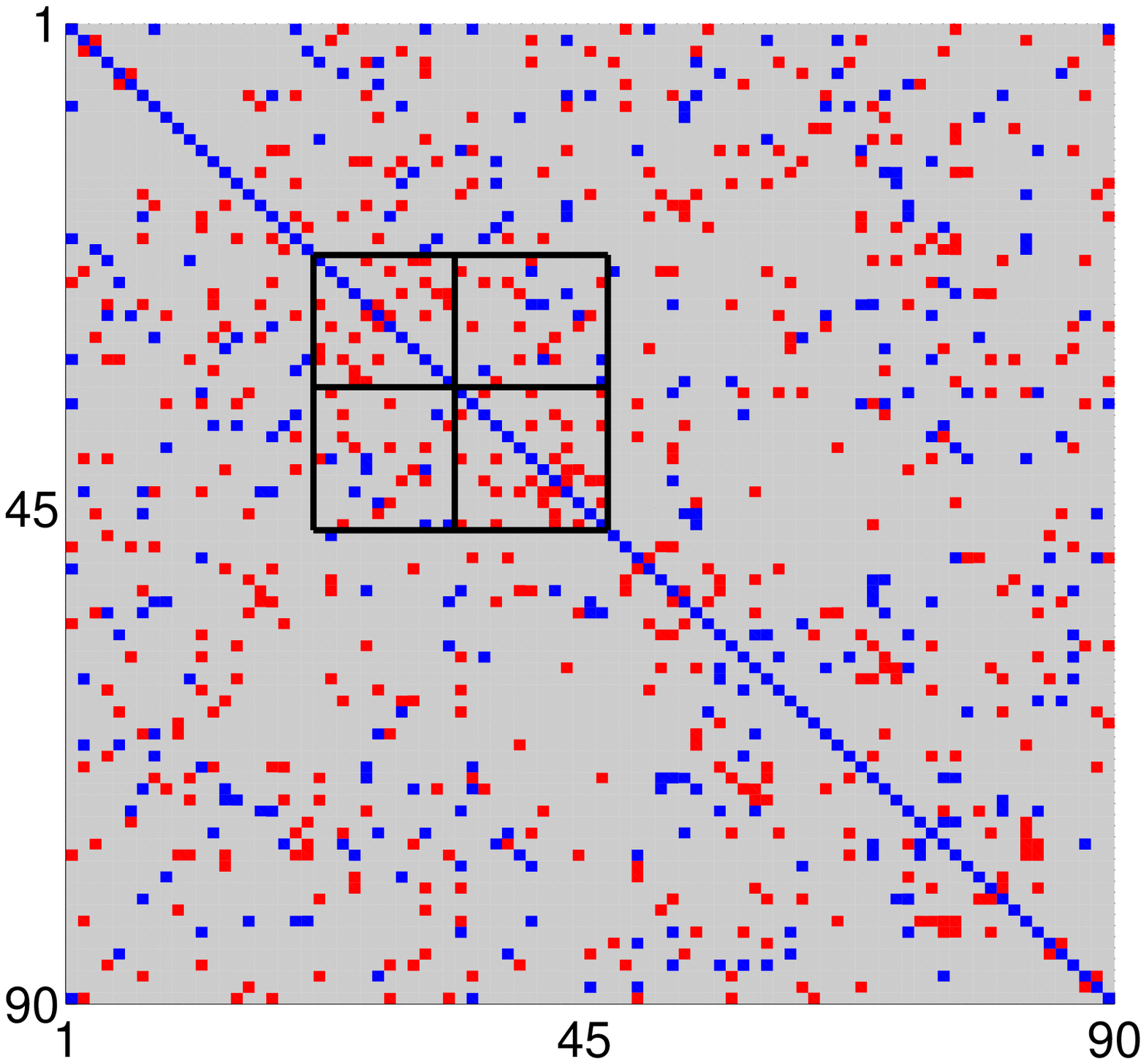}}&
	{\includegraphics[width=0.3\textwidth]{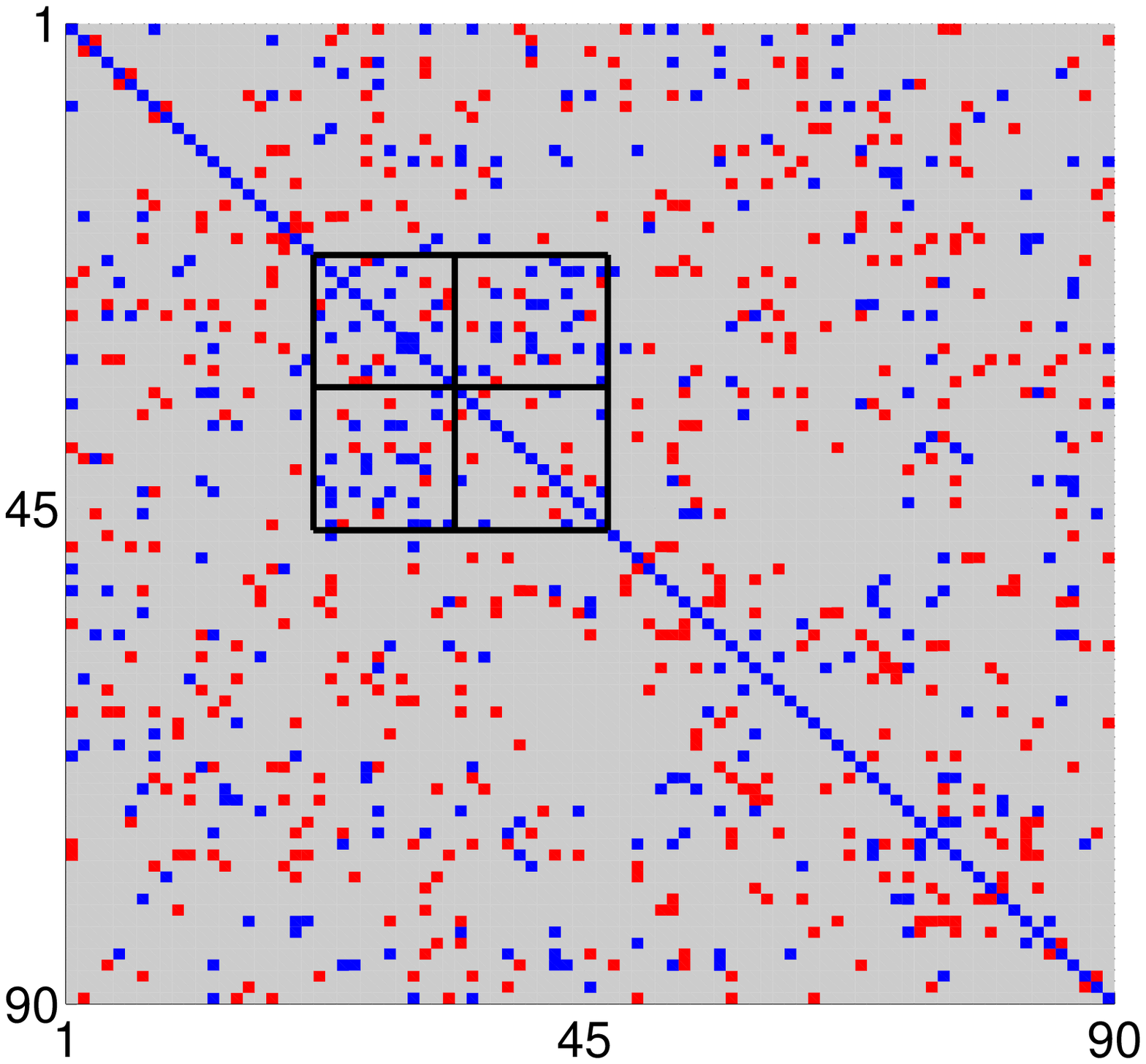}}&
		{\includegraphics[width=0.3\textwidth]{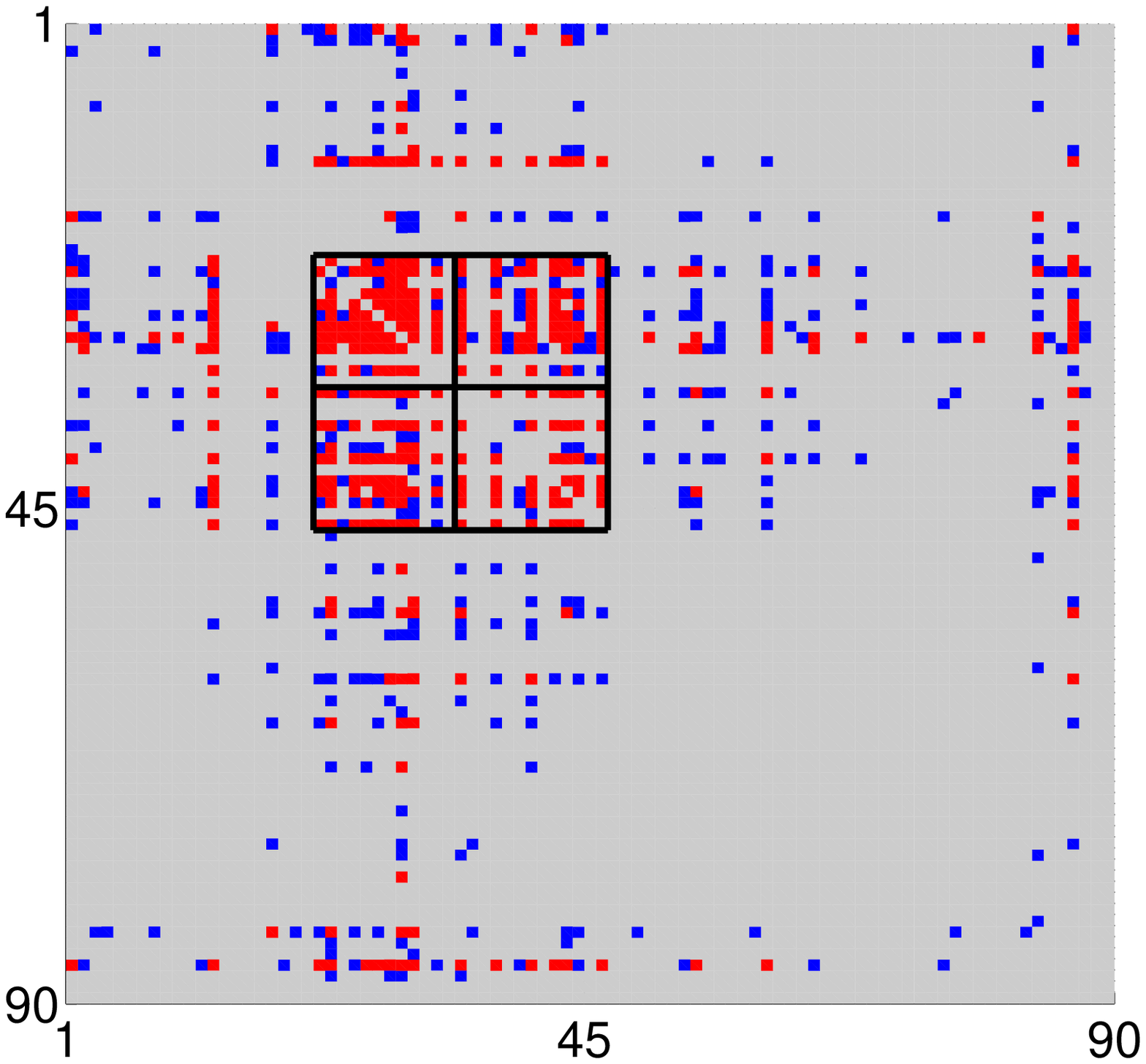}}\\
	\vspace{-12mm}\\
		{\footnotesize \hspace{-4mm}$PSC_{\bm\omega_{n}^T}(\theta=0)$} & \hspace{-4mm} {\footnotesize $PSC_{\bm\xi_{\omega;n}^{T}}(\theta=0)$} 
		& \hspace{-4mm}{\footnotesize$\vert PSC_{\bm\omega_{n}^T}(\theta=0)-PSC_{\bm\xi_{\omega;n}^{T}}(\theta=0)\vert $}\\
	\end{tabular}
		\vspace{-2mm}\\
	\end{center}
		\caption{\textnormal{$PSC$s at frequency $\theta=0$.~Left and middle panels: weights in absolute values below the~90$^{\mbox{\tiny {th}}}$ percentile in grey, weights above the~90$^{\mbox{\tiny {th}}}$ percentile in red, and below the~10$^{\mbox{\tiny{th}}}$ percentile in blue. Right panel: weights below the~90$^{\mbox{\tiny {th}}}$ percentile in grey, between the~90$^{\mbox{\tiny {th}}}$ and~95$^{\mbox{\tiny {th}}}$ percentiles in blue, and  above the~95$^{\mbox{\tiny {th}}}$ percentile in red.}\vspace{-3mm}} \label{fig:psc2}
\end{figure}

The two $PSC$s are shown in the first two panels of Figure \ref{fig:psc2}, while in the rightmost panel we show the absolute values of their differences. Inspection of these results  clearly indicates  that (i) after removal of the market shock, the idiosyncratic component still contains important dependencies, and (ii) an important benefit of our ``factor plus sparse VAR'' approach is to uncover the hidden dependencies between and within the Financial and Energy sectors. Moreover, when repeating the same analysis at other frequencies (e.g.,~$\theta=\pi/2,\  \pi$), no significant difference emerges between the two $PSC$s;  the benefits of our approach thus are relevant mainly in the long run. 

Similar conclusions can be derived by considering spectral densities (at different frequencies) and the squared partial spectral coherence (averaged over all frequencies), a measure which is non-zero if and only if two series are uncorrelated at all leads and lags after taking into account the (linear) effects of all other series in the panel. Results are in the Web-based supporting materials.
\begin{table}[t!]
\begin{center}
\footnotesize{
\begin{tabular}{l c c c c c |c}
\hline
\hline
\multicolumn{1}{r}{\footnotesize percentiles}& 50$^{\mbox{\tiny {th}}}$&	90$^{\mbox{\tiny {th}}}$&	95$^{\mbox{\tiny {th}}}$&	97.5$^{\mbox{\tiny {th}}}$&	99$^{\mbox{\tiny {th}}}$& \footnotesize{max}\\
\hline
\footnotesize $\vert PSC_{\bm\omega_{n}^T}(\theta=0)\vert$&0.06	&0.16	&0.18	&0.22	&0.25 & 0.35\\
 \hline
 \footnotesize $\vert PSC_{\bm\xi_{\omega;n}^{T}}(\theta=0)\vert$& 0.06&	0.16&	0.19&	0.22&	0.25 & 0.34\\
  \hline
 \footnotesize $\vert PSC_{\bm\omega_{n}^T}(\theta=0)-PSC_{\bm\xi_{\omega;n}^{T}}(\theta=0)\vert$& 0.02&	0.06&	0.08&	0.12&	0.15& 0.24\\
\hline
\hline
\end{tabular}
\vspace{-2mm}
}
\end{center}
\caption{\textnormal{Distribution of absolute value of $PSC$s' entries.}}\label{tab:psc}
\end{table}


\section{Conclusions} 
In this paper, we extend the study of interconnectedness of volatility panels initiated by \citet{Diebold:Yilmaz:2013} to the high-dimensional setting where a factor structure can be assumed for the data. We determine and quantify the different sources of variation driving a panel of volatilities of S\&P100 stocks 
 over the period 2000-2013. Our analysis highlights the key  role of the Financial sector, which appears to be particularly  important during the   Great Financial Crisis. Other sectors such as Energy, and in some cases also Health Care, seem to have an important role too. Moreover, we show that, contrary to a direct sparse VAR approach, our ``factor plus sparse VAR''  method can unveil crucial  inter-sectoral dependencies, which can be of tantamount importance for investors' decisions in the context of risk management. 

{\small   
\bibliography{BH_Biblio}   
\bibliographystyle{chicago}
}

\end{document}